\documentclass[noshowpacs,prb,eqsecnumbers,superscriptaddress,twocolumn]{revtex4}
\usepackage{amsmath}
\usepackage{mathtools}
\usepackage{epsfig,color}

\usepackage{graphicx}
\graphicspath{{images/}}
\usepackage{dcolumn}
\usepackage{bm}
\usepackage{hyperref}
\usepackage{cleveref}
\usepackage[english]{babel}

\newcommand{\beq}{\begin{equation}}
\newcommand{\be}{\begin{equation}}
\newcommand{\eeq}{\end{equation}}
\newcommand{\ee}{\end{equation}}
\newcommand{\bea}{\begin{eqnarray}}
\newcommand{\eea}{\end{eqnarray}}
\newcommand{\bwt}{\begin{widetext}}
\newcommand{\ewt}{\end{widetext}}

\begin{document}

\title{Multicritical Fermi surface topological transitions}

\author{Dmitry V. Efremov}
\thanks{These two authors contributed equally.}
\affiliation{Department of Physics, Loughborough University, Loughborough LE11 3TU, UK.}

\author{Alex Shtyk}
\thanks{These two authors contributed equally.}
\affiliation{Department of Physics, Harvard University, Cambridge, MA 02138, USA.}

\author{Andreas W. Rost}
\email{a.rost@st-andrews.ac.uk}
\affiliation{School of Physics and Astronomy, University of St Andrews, UK.}

\author{Claudio Chamon}
\affiliation{Department of Physics, Boston University, Boston, MA, 02215, USA.}

\author{Andrew P. Mackenzie}
\affiliation{School of Physics and Astronomy, University of St Andrews, UK.}
\affiliation{Max Planck Institute for Chemical Physics of Solids, Noethnitzer Strasse 40, 01187
Dresden, Germany.}

\author{Joseph J. Betouras}
\email{J.Betouras@lboro.ac.uk}
\affiliation{Department of Physics, Loughborough University, Loughborough LE11 3TU, UK.}

\date{\today}

\begin{abstract}

A wide variety of complex phases in quantum materials are driven by electron-electron interactions, which are enhanced through density of states peaks. A well known example occurs at van Hove singularities where
the Fermi surface undergoes a topological transition. Here we show that higher order singularities, where multiple disconnected leaves of Fermi surface touch all at once, naturally occur at points of high symmetry in the Brillouin zone. 
Such multicritical singularities can lead to stronger divergences in the density of states than canonical van Hove singularities, and critically boost the formation of complex quantum phases via interactions. As a concrete example of the power of these Fermi surface topological transitions, we demonstrate how they can be used in the analysis of experimental data on Sr$_3$Ru$_2$O$_7$. Understanding the related mechanisms opens up new avenues in material design of complex quantum phases.
\end{abstract}

\maketitle

%\twocolumn

{\it Introduction.} The properties of unconventional phases 
%density waves 
in quantum materials are generally connected to features of the electronic band structure. For example in density waves, characteristic wave vectors of emergent order parameters can often be related to nesting-type features of the underlying Fermi surface (FS) as discussed for e.g. iron pnictides \cite{Dai}, organics \cite{Monceau}, and transition metal dichalcogenides \cite{Ausloos}. Yet these nesting features in themselves usually cannot account for the observed thermodynamic stability of such correlated quantum phases. Intriguingly in a range of these materials the band structure hosts energetically close-by singularities in the density of states $\nu$ (DOS), which have been conjectured often to be crucial ingredients stabilising the emergent phases.

\begin{figure}[t ]
	\center{\includegraphics[width=1\linewidth]{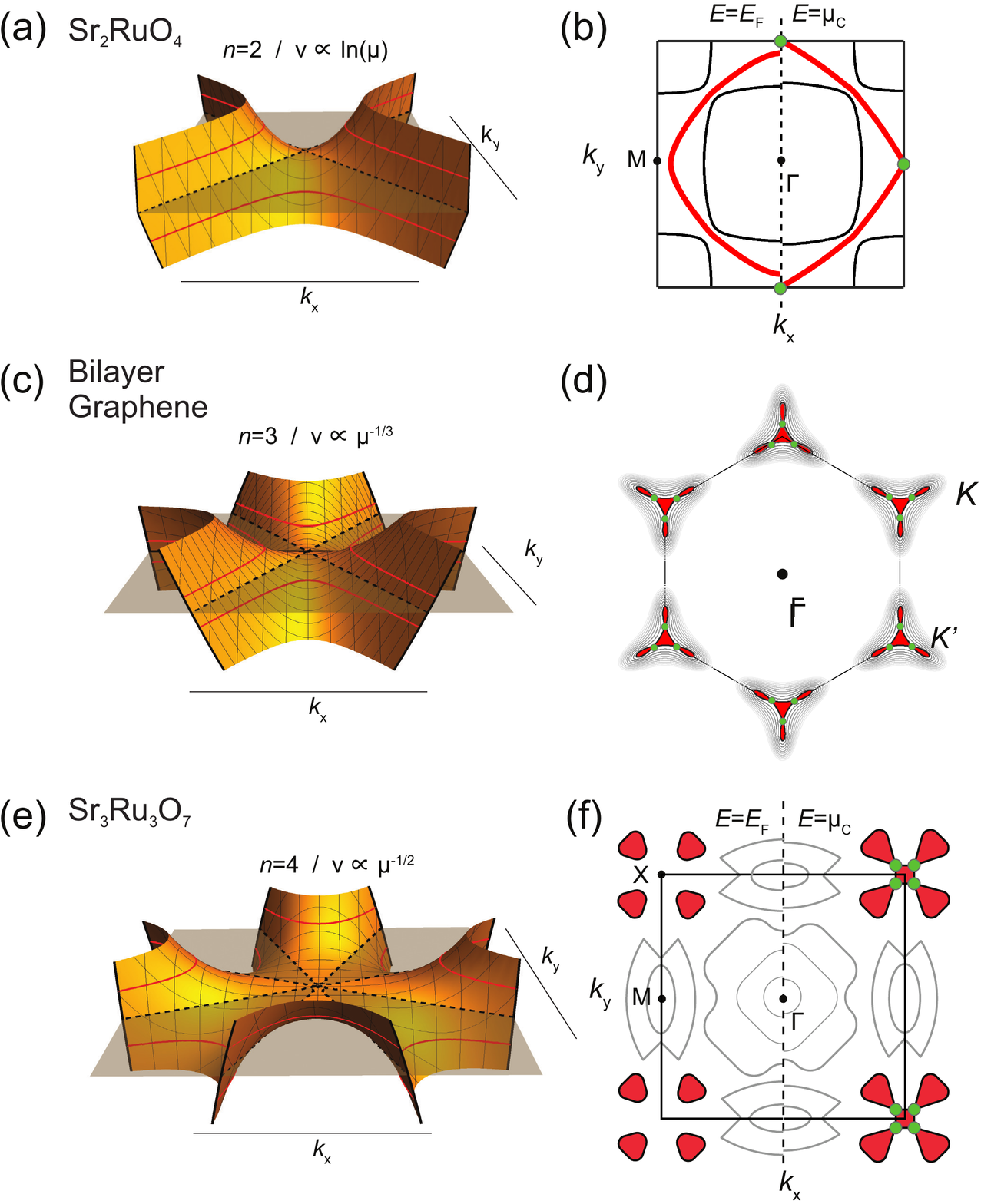}} 
	\caption{(a) The $n$=2 2D van Hove singularity in the form of a saddle point. The DOS $\nu$ diverges as $\ln\mu$. Red lines indicate FSs above and below the singularity. The critical FS is shown by the dotted black line. (b) This singularity occurs for example very close to the Fermi energy in the case of Sr$_2$RuO$_4$ (green marker for the critical chemical potential) (c) The $n$=3 singularity that can occur at 3-fold symmetric points. (d) The band structure / FS of bilayer graphene that is very close to such a $n=3$ monkey saddle singularity. (e) The $n=4$ singularity at a 4-fold symmetric point. (f) Schematic of the quasi-2D FS of Sr$_3$Ru$_2$O$_7$ in the $k_z$=0 plane at the Fermi energy (left hand side) and at $\mu_\textrm{C}$ (right side). The crucial bands that are close to the $n=4$ multicritical point are highlighted in red. The central pocket is a small perturbation (see main text). In order to emphasise the characteristic clover leaf FS we show an extended $k$-space picture beyond the BZ boundaries.}
	\label{fig:structure}
\end{figure}

Singularities in the DOS occur naturally at FS topological Lifshitz transitions (LT). A prominent example is the van Hove singularity (vHs) formed at a saddle point in the energy-momentum dispersion (see fig.~\ref{fig:structure}a). A two-dimensional (2D) vHs has a relatively weak logarithmic divergence in the DOS but is known to lead to a wealth of phenomena such as ferromagnetism driven by the Stoner mechanism (see eg. Ref. \cite{Rhodes}). An important point is that the thermodynamic stability of the emergent phases depends on the magnitude of the singularity as well as its shape \cite{Rhodes} (i.e. gradient and curvature).
% with the instability criterion given by $\nu_F\nu_F^{''}>(\nu_F^{'})^2$. 
As a consequence stronger power law divergences can have a much more dramatic impact on the formation of complex ordered phases than the weaker vHs.  

The identification of these singularities in correlated quantum materials is an important first step to understand their properties, given that many experimental quantities, with puzzling dependencies on the external probes, can be explained in a natural way.
Here, we explore the consequences of a generalisation of these concepts to multicritical topological transitions where multiple disjoint parts of a FS merge and demonstrate the power of the singularities in explaining properties of the correlated material  Sr$_3$Ru$_2$O$_7$.
%, as happens e.g. in some exceptional iron-based high temperature superconductors \cite{Reid} and biased bilayer graphene \cite{Shtyk}. 

Multicritical FS topological transitions naturally occur at points of high crystal symmetry, where the number $n$ of FS components merging depends on the particular symmetry. In fig.~\ref{fig:structure}a-c we illustrate the symmetries associated with the $n=2$ (vHs), 3 and 4 cases in 2D. When the singularity occurs at an edge of a Brillouin zone (BZ) there are generically two pieces ($n=2$) of the FS that join at the singularity, as depicted in fig.~\ref{fig:structure}a. At the critical energy (dotted line) there is a topology change in the FS structure with $n=2$ FSs touching and reconnecting. At the corner of a hexagonal BZ, three FSs or leaves ($n=3$) can join at the singularity, as happens in e.g. biased bilayer graphene \cite{Shtyk}, (fig.~\ref{fig:structure}b). In the square lattice, four leaves ($n=4$) can meet at the $X$-point in the corner of the BZ (fig.~\ref{fig:structure}c). At these high symmetry points higher order terms in the dispersion can become relevant, critically changing the divergence of the DOS, e.g. from a logarithmic to a stronger power-law divergence in the $n$=4 case discussed below.

%In a second step we will demonstrate how a multiple Lifshitz transition point can act as a crucial boost, enhancing density wave formation. By providing a strong feedback mechanism already at lowest order of perturbation theory a generic enhancement of pre-exisiting peaks in the general polarisability due to e.g. Fermi surface nesting can be achieved. This mechanism can ultimately drive the formation of complex density waves by Fermi surface nesting which would not be stable if only the bare susceptibilities were considered. We would like to emphasis that this is a purely electron-electron interaction driven mechanism of phase formation.
An explicit illustration of the experimental effects of these concepts is provided by examining the physics of the layered perovskite Sr$_3$Ru$_2$O$_7$, which has been intensely studied because of its unusual magnetic and transport properties \cite{Mackenzie}. We identify a strong power-law singularity at a four-fold symmetric point in momentum space (fig.~\ref{fig:structure}f) and demonstrate how, in conjunction with other features of the FS and electron interactions, this multicritical feature is pivotal for the physical properties of this material, explaining several previously perplexing characteristics. This is the first tangible demonstration in an existing layered material of the effects of such multicritical LTs (MLTs).

{\it Effective dispersion relation and analysis for n=4.} The energy-momentum dispersion relation in the vicinity of a $n=4$ MLT can be approximated by the simple expression
\begin{equation}
\label{toy}
	\varepsilon(\vec{k})=
	\begin{cases}
ak^2 + k^4\cos4\varphi-\mu ,
\\

a(k_x^2+k_y^2)+(k_x^4 - 6 k_x^2k_y^2+k_y^4)-\mu
\\
\end{cases}
\end{equation}
%\end{widetext}
where the first line is a representation in polar coordinates (with $\varphi$ being the azimuthal angle) and the second being a representation in Cartesian coordinates.

\begin{figure}[t]
	\center{\includegraphics[width=1\linewidth]{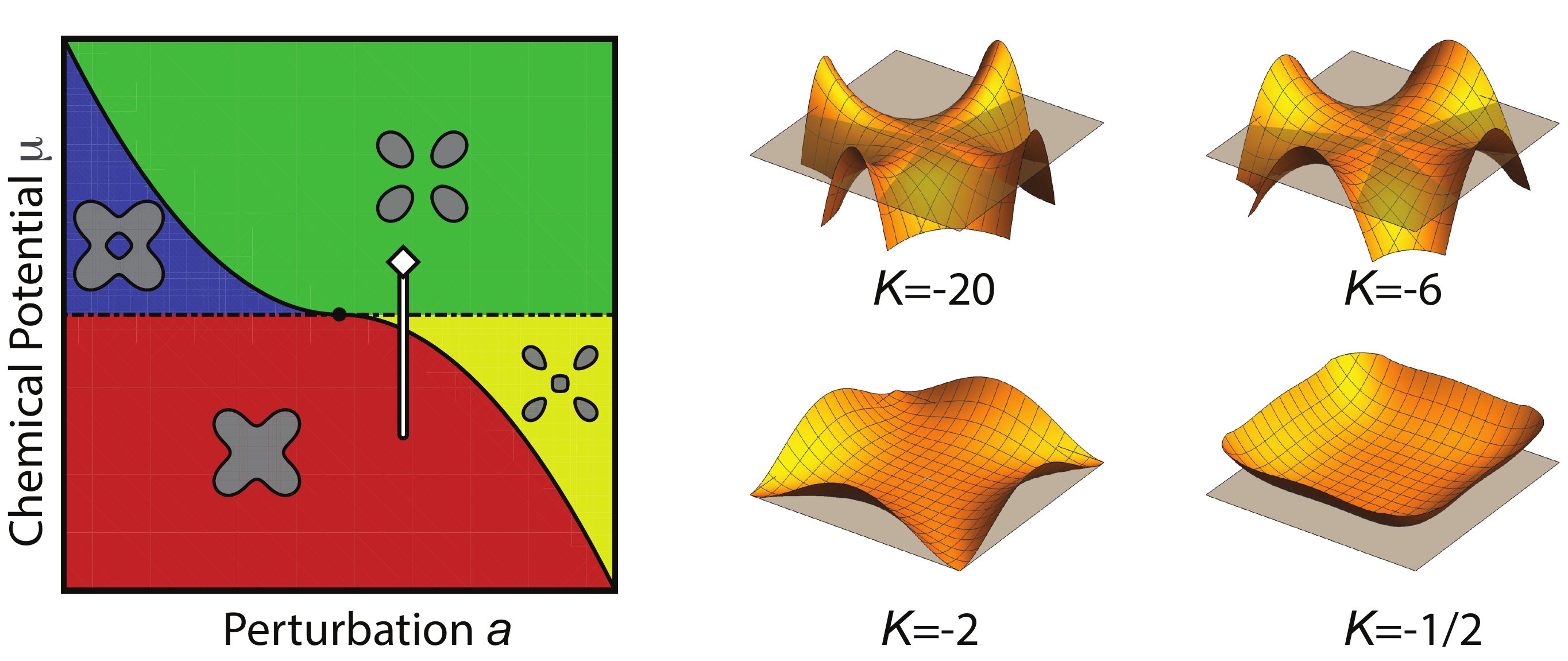}} 
	\caption{Schematic phase diagram of the relevant hole-like version of the effective dispersion relation (\ref{toy}). In Sr$_3$Ru$_2$O$_7$ there is no control over the parameter $a>0$, so that only red, yellow and green phases are accessible. There are two lines of LTs, a dashed-dotted line of a band edge type and a solid line corresponding to the transition of the neck-narrowing type. The $X_9$ singularity is located at the crossing of these two lines. The white line schematically shows the location of  Sr$_3$Ru$_2$O$_7$ within this phase diagram with the diamond marking the location in zero field. (Right side)  The three dimensional surfaces above are electron dispersions $\varepsilon=\varepsilon(k_x,k_y)$ in the vicinity of the singularity. The grey horizontal plane represents the critical energy of the singularity $\varepsilon=0$.}
	\label{fig:phase_diagram}
\end{figure}

For small non-zero positive $a$ this dispersion exhibits two LTs as we change the chemical potential $\mu$ (see Fig.~\ref{fig:phase_diagram}). At smaller values of $\mu$ one large hole-like FS exists. At the critical chemical potential four vHs appear at a FS topological transition where a new center pocket is created. As the chemical potential is increased further, the FS undergoes a second LT, of the band edge type, with the vanishing of the centre pocket.

If the dispersion relation can be tuned closer to $a=0$, then the singularity is approached, the vHs merge with the minimum of the central electron pocket to form a 4th order saddle.
\begin{equation}
4\times \underbrace{(k_x^2-k_y^2)}_{\mathclap{\text{vH saddle}}}+1\times \underbrace{k^2}_{\mathclap{\text{e/h pocket}}}
\longleftrightarrow
\underbrace{k^4\cos4\varphi}_{\text{4$^\text{th}$ order saddle}},
\end{equation}
%
%It is essentially a bifurcation of the 4th order saddle into four van Hove saddles (and a band minimum at the center).

The singularity can be viewed as a Lifshitz multicritical point, as it appears at the crossing of two LTs and sits at the border of four different topological phases, see Fig.~\ref{fig:phase_diagram}. Such behavior can be described within the framework of singularity theory~\cite{Arnold} by a symmetry-restricted unimodal parabolic singularity $X_9$ in the electron dispersion $\varepsilon(\vec{k})$. 
%In Cartesian coordinates, the electron dispersion in the vicinity of the singularity can be captured by the dispersion
%\begin{equation}
%\label{eq:x9}
%\varepsilon(\vec{k})=(k_x^4+Kk_x^2k_y^2+k_y^4)+a(k_x^2+k_y^2)-\mu.
%\end{equation}
The core of the singularity is the 4$^\mathrm{th}$ order terms, generalised to $k_x^4+Kk_x^2k_y^2+k_y^4$ in Cartesian coordinates.
%, where we had rescaled momenta to eliminate the coefficient in front of them
This term is the germ of the singularity, while the remaining terms $(a(k_x^2+k_y^2)-\mu)$ represent the perturbation unfolding the singularity. Unlike simpler singularities, $X_9$ forms a whole family of singularities parametrized by the modulus $K$. While a generic singularity from the $X_9$ family has a co-dimension eight, one modulus and seven control parameters, the presence of the lattice symmetry greatly simplifies the situation leaving only the modulus $K$ and two control parameters $a, \mu$.  The consequences of the singularity on physical properties are the same for the whole range $K<-2$.
The value of the modulus $K=-6$ is special as it corresponds to electron-like and hole-like sections of the same width, a property that is confirmed in the DFT calculation. This implies existence of an additional symmetry in the system, a superposition of the particle-hole transformation $\varepsilon\leftrightarrow-\varepsilon$ and rotation by an angle $\pi/4$. If we increase the value to $K=-2$ the system reaches a critical point and the saddle disappears, leaving a singular $\propto k^4$ electronic pocket. All values of the modulus $K<-2$ lead to the same topological features. 

%At this special value the low-energy dispersion acquires an additional symmetry that is composed from the particle-hole transformation and rotation by an angle $\pi/4$.
%with dispersion acquiring especially simple form in polar coordinates, which is Eq.(\ref{toy}).

%$(k, \varphi)$:
%\begin{equation}
%\label{toy}
%\left.\varepsilon(\vec{k})\right|_{K=-6}=k^4\cos4\varphi+ak^2-\mu.
%\end{equation}
%
%The last term in the Eq.(\ref{toy}) is irrelevant for the singularity. It's purpose is to close the Fermi surface at larger momenta. 

In the relevant parameter regime $K<-2$ the DOS of this dispersion has a critical $\propto|\mu|^{-1/2}$ scaling for $a=0$ and can be summarized as 
\begin{equation}
\label{eq:div}
\nu(\mu)\propto
\begin{cases}
\left|\mu-\mu_c\right|^{-1/2}, & \left|\mu\right|\gg \mu_c
\\
\ln \dfrac{\mu_c}{\left|\mu-\mu_c\right|}, & \left|\mu-\mu_c\right|\ll \mu_c,
\end{cases}
\end{equation}
where the critical value of the chemical potential $\mu_c=a|a|/4$ (more details in the SM).

{\it Experimental consequences of the singularities in correlated systems: Sr$_3$Ru$_2$O$_7$.}
There are profound consequences of this singularity for Sr$_3$Ru$_2$O$_7$, a material of wide interest due to the observed phenomena in the vicinity of a metamagnetic quantum critical end point (QCEP) at $H_c$=8 T for fields parallel to the crystallographic $c$-axis \cite{GrigeraScience, Grigera}.  %  The strong electron correlations related to this quantum critical end point are evident in thermodynamic properties. %In contrast, hydrostatic pressure weakens the magnetism\cite{Chiao}, as qualitatively expected in a Stoner picture in which peaks in the DOS near the Fermi level enhance charge and spin susceptibilities.

\begin{figure}[b]
	\center{\includegraphics[width=0.57\linewidth]{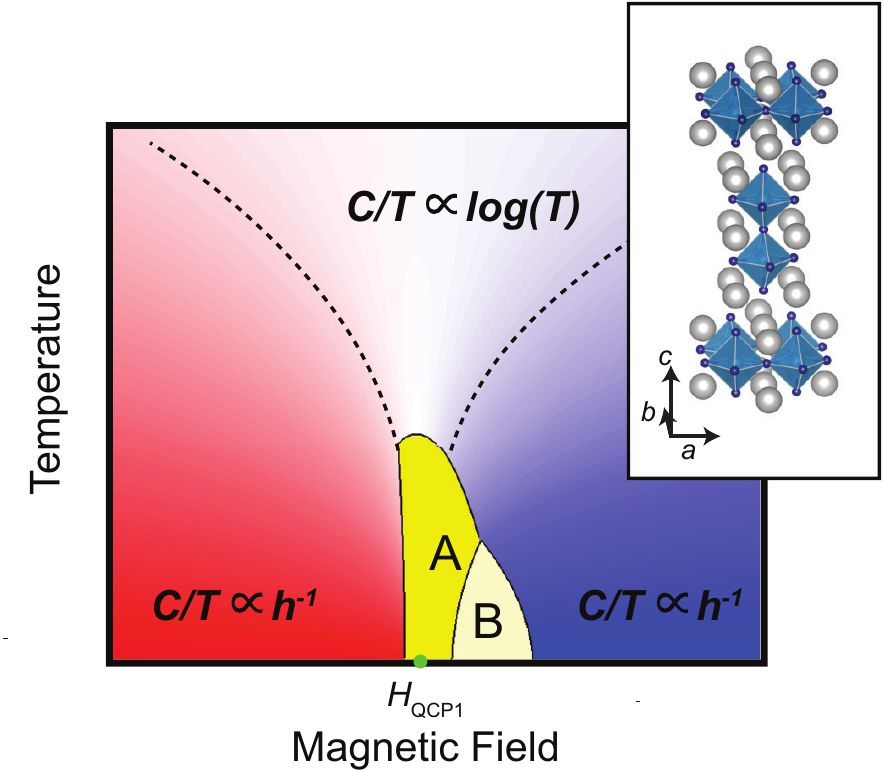}} 
	\caption{Schematic phase diagram of Sr$_3$Ru$_2$O$_7$ together with the crystal structure shown in the inset (Ref. \onlinecite{Mackenzie} for more details).}
	\label{fig:327PD}
\end{figure}

In Fig.~\ref{fig:327PD} the emergent phase diagram is schematically shown highlighting several features relevant for our discussion (for a review \cite{Mackenzie}). Approaching $H_c$ as a function of $T$, a logarithmic divergence in specific heat divided by temperature $C/T$  is observed \cite{Perry, Rost2}. The approach to $H_c$ as a function of magnetic field in the Fermi liquid (FL) regime is characterised by a singular contribution  to $C/T$. Careful analysis \cite{Rost} reveals a power law divergence of $C/T$ as a function of reduced field $h=(H-H_C)/H_C$ with an unexpected exponent of (-1). It has been suggested that the singularities in $C/T$ as a function of field or temperature are consistent with a 2D QCEP point within the canonical description of quantum criticality \cite{Gegenwart}. The expected exponent within this theoretical framework is -1/3 and in general has to be fixed in any fit of this model \cite{Sun}. The observed exponent of (-1) in an assumption-free power law fit to $C/T$ therefore posed important theoretical questions.

At low temperature, access to the QCEP is preempted by an unusual set of emergent phases (labelled A, B in Fig.~\ref{fig:structure}d) \cite{Borzi, Lester}.  Neutron scattering measurements \cite{Lester} revealed an incommensurate magnetic order with a wave vector $\vec{Q}^A= (\pm 0.233,  0,  0)$/$(0, \pm 0.233, 0)$ and  $\vec{Q}^B= (\pm 0.218,  0,  0)$/$(0, \pm 0.218, 0)$ respectively. 

\begin{figure}[t!]
	\center{\includegraphics[width=0.9\linewidth]{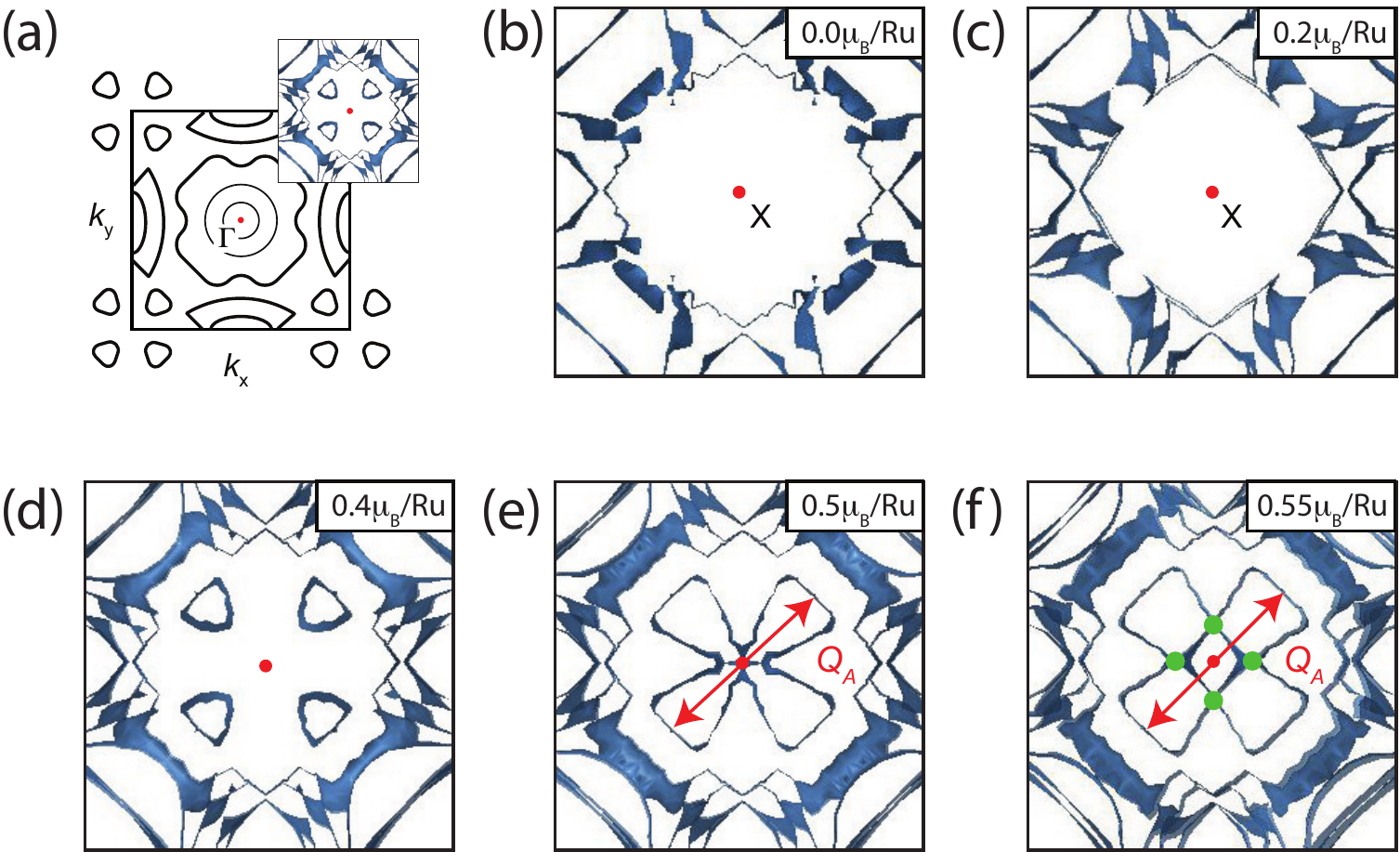}} 
	\caption{Result of the DFT calculation. (a) Here we show the schematic FS structure from Fig.~\ref{fig:structure}e  together with the $k_z$-projected DFT calculation to aid orientation. (b-f) Projected FSs for values of magnetization $\mu$=0.0, 0.2,0.4,0.5,0.55 $\mu_B$ per unit cell as calculated by the DFT and centered at the X point. The topological transitions are evident for the value of magnetization close to 0.5 $\mu_B$  and 0.55 $\mu_B$ per unit cell.
	The red arrow in (e,f) connects the two nested parts of the characteristic clover leaf structure giving rise to the density wave in the A-phase.}
	\label{fig:FS_magnetization}
\end{figure}

{\it Fermi surface as calculated by Density Functional Theory (DFT).}~~~
To understand the origin of the singularity in the DOS we performed DFT calculations \cite{Supplemental, Koepernik, URL}.  The calculated band structure for zero magnetization agrees broadly with ARPES data \cite{Tamai, Allan} (Fig.~\ref{fig:FS_magnetization}a). While the chemical potential is slightly higher than observed experimentally, this does not affect the main conclusions drawn here. We therefore consider the evolution of the DOS with increasing magnetic moment per unit cell (see fig.~\ref{fig:FS_magnetization}), as a convenient way to model the effects of an applied magnetic field. By increasing the magnetic moment 
%by 0.1 $\mu_B$/Ru 
a LT is observed at the X-point at a magnetization of around 0.5 $\mu_B$/Ru. This LT dominates the DOS and the thermodynamic properties. In order to identify the essential requirements generating the key FS features we derive a quasi-2D tight binding model based on the Ru 4d orbitals (see Supplemental Material (SM) \cite{Supplemental}) relevant at the Fermi energy \cite{Puetter} and adjusted to accurately describe the relevant part of the ARPES data. 
The resulting FS is shown in fig.~\ref{fig:structure}f. 
%The multicritical LT effectively originates from (i) 2D hopping terms on a square lattice (predominantly within the $d_{xy}$ orbitals) in combination with (ii) the doubling of the unit cell due to RuO-octahedra rotation (which also mixes in $e_g$ orbitals as a secondary effect) and (iii) bilayer splitting. 
A careful study of the band structure reveals that it is well described by an effective dispersion given in eq. (\ref{toy}). The DFT calculations suggest a value of $K$ close to $K=-6$, implying an effective power law divergence (eq. (\ref{eq:div})). 

Qualitatively similar conclusions can be drawn from the tight binding model which unveils the microscopic origin of the MLT point (see SM). In the aristotype structure without RuO$_2$ octahedra rotations the band structure exhibits a $n=2$ vHs at an $M$-point of the BZ. Counterintuitively the rotations, while lowering the crystal symmetry, reconstructs the BZ such that the singularity is transformed into an $n=4$ MLT point at an $X$- point in the new BZ. This is an important guiding principle how to stabilise such LT.

The characteristic clover leaf structure of the FSs in the vicinity of the LT naturally gives rise to strong nesting of the edges of the $\gamma$-bands which, in combination with the MLT, helps generate the SDW. It is important to note that the nested FS parts have a distinct orbital character ($d_{xz/yz}$) from those FS parts that create the MLT ($d_{xy}$). The value of the nesting vector ${\vec{Q}}=(\pm 0.23, 0, 0)$ or $(0, \pm 0.23, 0)$ is effectively that observed experimentally in phase A. These nested parts of the FS determine the wavelength of the observed SDW (see also \cite{Lester}) within our theoretical model. %There is no particularly large DOS associated with the nesting in itself and this feature alone cannot stabilise the SDW - this is only possible by interacting with the energetically close by (but from a band structure/orbital character point of view effectively independent) MLT. 

{\it Magnetic field approach to criticality}. The MLT has a profound effect on the specific heat $C_v$ as a function of the magnetic field ($C_v \propto |H-H_c|^{-1}$). The power-law divergence of the DOS as a function of energy leads to the divergence of $C_v$ with the field \cite{Rost}. $C_v$ is determined by the value $\nu(\epsilon_F)$ of the DOS at the Fermi level: $C_v=\frac{\pi}{2} k_B T \nu(\epsilon_F)$. Near the singularity $\nu(\epsilon)\propto|{\epsilon-{\epsilon}_c}|^{-1/2}$, where ${\epsilon}_c$ is the location of the singularity. Including magnetic field to lowest order, the DOS is $\nu(\epsilon, H)=\frac{1}{2} \left( \nu(\epsilon + g \mu_B H) + \nu(\epsilon - g \mu_B H) \right) $. Charge conservation requires the Fermi energy to shift non-linearly with $H$:
\begin{eqnarray}
\nonumber
\left[\epsilon_F(H) - \epsilon_c + g \mu_B H\right]^{1/2} + \left[\epsilon_F(H) - \epsilon_c - g \mu_B H\right]^{1/2} \\
\label{app:central}
= 2 \left[\epsilon_F(H=0) - \epsilon_c\right]^{1/2}
%\end{flalign}
\end{eqnarray}
\newline
At $H_c$, $\epsilon_F(H_c)  = \epsilon_c + g \mu_B H_c$ as the singularity is within the minority band. Then $\epsilon_F(0) - \epsilon_c = \frac{1}{2}  g \mu_B H_c$ and for $H$ near $H_c$ Eq.(\ref{app:central}) reads $\epsilon_F(H) = \epsilon_c + g \mu_B \frac{H^2 + H_c^2}{2 H_c}$.
Therefore, $\nu(\epsilon_F(H)) \propto 1/|H-H_c|$  and the specific heat as well as the entropy is proportional to $1/|H-H_c|$. This explains the experimental data and is a direct fingerprint of the existence and the importance of the $n=4$MLT in Sr$_3$Ru$_2$O$_7$.

%Close to $H_c$ the system is susceptible to any perturbation, because of the divergent DOS. In the vicinity of this field two main phenomena have been observed: (i) the metamagnetic transition and (ii) the formation of SDW. To the accuracy of the experimental resolution, both transitions happen almost at the same value of the magnetic field. The $\gamma$ bands are responsible for both, as we have already seen, but the important observation is that different parts of the $\gamma$ bands are responsible for each of them; the centre of the $\gamma$ bands as the one exhibiting the power-law singularity in DOS is responsible for the meta-magnetic transition, while the edges of the $\gamma$ bands (with a non-singular DOS) are responsible for the formation of the SDW. 

%{\bf {\it Phase Formation.}} 
{\it Phase Formation}. As explained, there is significant nesting along $\vec{Q}^A$ in the $\gamma$ band giving rise to a susceptibility to SDW formation. As the SDW and the MLT originate from different orbitally orthogonal parts of the FS, at tree-level in a Renormalization Group (RG) sense the two processes can be treated as decoupled, At higher RG order this is no longer true and the DOS singularity influences the thermodynamic stability of the SDW. 
The mechanism of this particular LT as described above, involves the creation of a pocket at the X-point. An effect of interactions is that this pocket formation is a first order transition with a jump to a higher total number of fermions in the relevant bands \cite{Betouras2}. This is consistent with experimental observations upon entering the A-phase as a function of field. The additional pocket, in combination with the MLT point, provides additional FS degrees of freedom. This leads to the counterintuitive result of the high-field A-phase having a higher entropy than the low field FL phase, as is established experimentally \cite{Rost}.
%As a result, the entropy jumps to a higher value at $H_c$ when entering the A-phase, reflecting that change. 
%This explains the asymmetry between the two curves $\propto 1/|H-H_c|$ on the left and right of $H_c$.
%The free energy close to $H_c$ due to the formation of the SDW is given in the SM, using scaling theory \cite{Zhu} with a dynamical exponent $z=2$. The coherence length diverges near the transition as $\xi \propto |H-H_c|^{- 1/2}$. The entropy increases near the SDW transition as shown in the SM. 
%
%The metamagnetic transition also is of first order as the free energy acquires non-analytic terms. For example, at $T=0$ and for the DOS corresponding to $a=0$ for simplicity, we find a term in the free energy proportional to $|\delta m|^{3/2}$ where $\delta m$ is the jump in magnetisation. This implies that a Ginzburg-Landau free energy expansion breaks down in the case of the multicritical LT. For completeness, in the Supplemental Information we present, using scaling theory \cite{Zhu}, the part of the free energy also due to the SDW formation, with an entropy that is logarithmically divergent at $H_c$. A more detailed treatment that couples the two phenomena is beyond the scope of the present work.

{\it Temperature approach to criticality}. The logarithmic divergence $C_v \propto T$ log$(1/T)$ (Fig.~\ref{fig:structure}d) as a function of temperature is clearly an effect of interactions and a sign of quantum criticality. As shown previously \cite{Betouras1}, the formation of a small pocket in the middle of a larger FS leads to the same result due to interactions. Alternatively, it can be thought as a consequence of the scattering of "light" electrons further from the singularity,  off "heavy" electrons \cite{Berg}.
%Within the theoretical framework outlined two factors contribute. 
In addition, a correlated 2D system with self-energy which is position-dependent ($k$-dependent) leads to the same behavior \cite{Sachdev, Andrey}. In the case of SDW formation, the self-energy correction is $\Sigma(\omega, \vec{k}) \propto \omega/k$ (with $k=k_{||}$), giving an effective mass $m^* = m [1 - {\partial \Sigma}/{\partial \omega} |_{\omega \rightarrow 0}]$. Therefore, given that $C_v \propto T \int m^* dk$, then naturally $C_v \propto T$log$(1/T)$. These conditions are fundamentally linked to the MLT, leading to qualitatively similar behaviour.

{\it Discussion}
The effects of simple LTs were explored in several classes of quantum materials (e.g \cite{Betouras1, Sherkunov, Sherkunov2, Ptok, Volovik}).
In this work, we have demonstrated how a MLT is formed at a high symmetry point in the BZ can lead to a wealth of unusual physical phenomena. This was illustrated concretely through the example of Sr$_3$Ru$_2$O$_7$, in which a MLT happens in the $\gamma$ bands at the X-point of the BZ, leading to a 4-leaf ($n=4$) FS. This is accompanied by a large peak in the DOS, describable by a divergent as inverse square-root in energy singularity. We showed how this power-law in combination with the emergent central pocket and enhanced spin-/charge-susceptibilities due to independent, orbitally orthogonal parts of the FS are consistent with a wide range of previously puzzling experimental data. 
The intriguing behavior of this well-characterised material is explained within the framework of higher singularities. For any $n=4$ singularity in quasi-2D materials the phase diagram presented in Fig. 2 is relevant and whenever $K<-2$ a regime of effective power-law divergence should appear.

%\textcolor{red}{
The example of Sr$_3$Ru$_2$O$_7$ demonstrates one important finding;  the FL parameters at zero-field are extremely robust against tuning over a much wider range than previously believed. A direct implication would be that quantum fluctuation corrections are not relevant over a wide region of magnetic fields in the FL regime of the phase diagram. For example, significant further band renormalisation might not be expected. This is not only surprising but should also lead to a careful reevaluation of other materials.
% including (i) singular contributions to $C/T$ approaching $H_C$, (ii) the formation of the SDW close to the metamagnetic quantum critical end point and (iii) entropy changes associated with entering the A-phase from low field.
%The nested parts do not involve the singularity (at the centre of the $\gamma$-bands), therefore contributes less to the thermodynamics, which is controlled by the singularity. 
While the singularity dominates the thermodynamics, the heavy quasiparticles only indirectly contributes to transport through the scattering of electrons from other parts of the FS off them. Therefore, there is a subtle interplay of the importance of each part of the FS to different experiments.
%Furthermore, the transition from the A-phase to B-phase, which has not been considered in this work, is most likely driven by interaction effects and the renormalisation of the nested vector due to quantum fluctuations.
We note that we only partially discussed key aspects of the phase diagram of Sr$_3$Ru$_2$O$_7$ where quantum fluctuations contribute significantly.
% (e.g. the $T$log$(1/T)$ behavior of the specific heat and overall renormalisation of the band structure as evidenced by the large Wilson ratio).

Sr$_3$Ru$_2$O$_7$ serves as a model system and guide to a whole range of material classes in which MLTs occur (see SM for a discussion on generic mechanisms in e.g. ruthenates \cite{Baumberger}, bilayer graphene \cite{Sherkunov2, Shtyk, Yuan, Cao2} and transition metal dichalcogenides \cite{King}).
One key lesson is that the non-trivial divergences in the DOS are a key driver in thermodynamically stabilising unconventional phases, originating from otherwise independent (orbitally orthogonal) parts of the FS. The achievement of this work is to identify the importance of the MLT and thereby help to disentangle the roles of the LT in the band structure on the one hand and (quantum) fluctuations and interactions on the other. In all quasi-2D materials with an $n=4$ singularity the phase diagram presented in this work is relevant and whenever $K<-2$ a regime of power-law divergence should appear.
The counterintuitive mechanism that turns a trivial $n=2$ vHs into a $n=4$ MLT by lowering the crystal symmetry, provides a new guiding principle how to stabilise complex quantum phases.

\begin{acknowledgments}
We thank F. Baumberger, C. Castelnovo, A. Chubukov, G. Goldstein, C. Hicks and Y. Sherkunov for helpful discussions.  The work was supported by the EPSRC grant No. EP/P002811/1 and Royal Society (JJB) and DOE Grant No. DE-FG02-06ER46316 (CC).
\end{acknowledgments}

\newpage

\begin{widetext}
{\center{$\;\;\;\;\;\;\;\;\;\;\;\;\;\;\;\;\;\;\;\;\;\;\;\;\;\;\;\;\;\;\;\;\;\;\;\;\;\;\;\;\;\;\;\;\;\;\;\;\;\;\;\;\;\;\;\;\;\;\;$SUPPLEMENTAL MATERIAL}}

\setcounter{equation}{0}
\setcounter{figure}{0}
\setcounter{table}{0}
\setcounter{page}{1}
\renewcommand{\theequation}{S\arabic{equation}}
\renewcommand{\thefigure}{S\arabic{figure}}
\renewcommand{\bibnumfmt}[1]{[S#1]}
\renewcommand{\citenumfont}[1]{S#1}

\section{Density of states in the vicinity of the multicritical point}
\label{appendix2}

As we discuss in the main text, the physics of the multicritical Lifshitz point can be described by the dispersion
\begin{equation}
	\varepsilon(\vec{p})=ap^2 + bp^4\cos4\varphi+cp^8-\mu.
\end{equation}
Below we assume that $a$ and $\mu$ take small values. In this case the $p^8$ term is not crucial for the analysis, as it only serves to close the Fermi surface away from the singularity, and it can be omitted safely,
\begin{equation}
\label{app:simplified}
	\varepsilon(\vec{p})\simeq ap^2 + bp^4\cos4\varphi-\mu.
\end{equation}
For definiteness we assume $a>0$ below. The symmetry of the dispersion (\ref{app:simplified}) implies the relation $\nu(-|a|,\mu)=\nu(|a|,-\mu)$. 

The DoS of \ref{app:simplified} is given by
\begin{equation}
\label{app:gen}
\nu(\mu) = \int\frac{d^2p}{(2\pi)^2}\delta(ap^2 + bp^4\cos4\varphi-\mu)
=
\int\frac{d\varphi dt}{8\pi^2}
\delta(at + bt^2\cos4\varphi-\mu)
=\frac{1}{4\pi^2a}D\left(\frac{4b\mu}{a^2}\right),
\end{equation}
where we made a substitution $t=p^2$ and the function $D(x)$ is an elliptic integral
\begin{equation}
	D(x) = \int_0^{2\pi}d\varphi\int_{0}^{\infty}dt\,\delta(2t + t^2\cos\varphi-x)
	=2\Re\int_{0}^{\infty}dt\frac{1}{\sqrt{t^4-(2t-x)^2}}.
\end{equation}

The DoS obtained above has a natural energy scale 
\begin{equation}
	\mu_c=a^2/4b.
\end{equation}

\subsection{Critical scaling at $|\mu|\gg\mu_c$}

The term $ap^2$ breaks the multicriticality, so that at large values of the chemical potential $|\mu|\gg a^2/4b$, when the quadratic term can be neglected, the dispersion reduces to the pure fourth-order saddle
\begin{equation}
	\varepsilon(\vec{p}) = bp^4\cos4\varphi-\mu
\end{equation}
with the critical scaling of the DoS $\nu(\mu)\propto|\mu|^{-1/2}$,
\begin{equation}
\begin{split}
	\nu(\mu) &= \int\frac{d^2p}{(2\pi)^2}\delta(bp^4\cos4\varphi-\mu)=
	\frac{1}{4\pi^2\sqrt{b|\mu|}}
	\int d^2k\,\delta(k^4\cos4\varphi-1)
	=\frac{K(1/2)}{4\sqrt{2}\pi^{2}}\frac{1}{\sqrt{b|\mu|}}\propto|\mu|^{-1/2},
\end{split}
\end{equation}
where $K(1/\sqrt{2})\approx1.85$ is a complete elliptic integral of the first kind.

\subsection{Jump at $\mu=0$}

As we approach the singularity from the region of negative chemical potential, the quadratic term leads to the formation of an electron pocket at the center. The electron pocket forms at $\mu=0$ and leads to a jump in the density of states,
\begin{equation}
\begin{split}
\nu(\mu=+0)-\nu(\mu=-0) &= \lim\limits_{\mu\rightarrow+0}\int\frac{d^2p}{(2\pi)^2}\delta(ap^2 + bp^4\cos4\varphi-\mu)-\lim\limits_{\mu\rightarrow-0}\int\frac{d^2p}{(2\pi)^2}\delta(ap^2 + bp^4\cos4\varphi-\mu)
\\
&=\lim\limits_{\mu\rightarrow+0}\int\frac{d^2p}{(2\pi)^2}\delta(ap^2-\mu)=\frac{1}{4\pi a}.
\end{split}
\end{equation}

\subsection{Van Hove singularity at $\mu=\mu_c$}
Finally, at the value of the chemical potential $\mu=\mu_c=a^2/4b$ the electron pocket formed at $\mu=0$ touches the four outer leaves of the Fermi surface via the formation of four saddle points, located at $p_c=\sqrt{a/2b},\,\cos4\varphi_c=-1$:
\begin{equation}
\begin{split}
	ap^2 + bp^4\cos4\varphi-\mu&\approx a(p_c+\Delta p)^2 - b(p_c+\Delta p)^4\left(1-\frac{(4\Delta\varphi)^2}{2}\right)-\mu
	\\
	&=
	(ap_c^2-bp_c^4-\mu) + (2ap_c-4bp_c^3)\Delta p+ (a-6bp_c^2)(\Delta p)^2 + 8bp_c^4(\Delta\varphi)^2
	\\
	&=(\mu-\mu_c)-2a(\Delta p)^2+8\mu_c(\Delta\varphi)^2.
\end{split}
\end{equation}

The saddle point result in the logarithmic divergence in the DoS:
\begin{equation}
\begin{split}
	\nu(\mu) &= 4\int\frac{d^2p}{(2\pi)^2}\delta((\mu-\mu_c)-2a(\Delta p)^2+8\mu_c(\Delta\varphi)^2)
	\\
	&=
	\frac{1}{2\sqrt{2}\pi^2a}\int dxdy\,
	\delta\left(\frac{\mu-\mu_c}{\mu_c}-x^2+y^2\right)
	\\
	&\simeq \frac{1}{\sqrt{2}\pi^2a}\log\frac{\mu_c}{|\mu-\mu_c|}.
\end{split}
\end{equation}

These results can be confirmed by studying the limits of the general expression (\ref{app:gen}).

\subsection{General expression (\ref{app:gen})} 

Using a substitution $z= (t-x)/t$ the elliptic integral $D(x)$ can be rewritten as
\begin{equation}
D(x) = 2\Re\int_{- x}^{\infty}\frac{dz}{\sqrt{[(|x|-1)+z^2][(|x|+1)-z^2]}}.
\end{equation}
Taking the real part of the integral above, it just reduces the integration to the region where the argument of the square root is positive. Depending on the value of $x$, the true domain of integration is
\begin{equation}
\begin{split}
	x>1:&\quad z\in(-1,\,\sqrt{1+\|x|}),
	\\
	0<x<1:&\quad z\in(-1,\,-\sqrt{1-|x|})\cup(\sqrt{1-|x|},\,\sqrt{1+|x|}),
	\\
	x<0:&\quad z\in(1,\,\sqrt{1+|x|}).
\end{split}
\end{equation}

Transforming the variable as $z = \sqrt{(|x|+1)}\cos \theta$ we bring the integral to the canonical form
\begin{equation}
	D(x)=\sqrt{\frac{2}{|x|}}\times
	\begin{cases}
		F(\pi-\varphi_1(x), k(x)), & x>1
		\\
		%F(\pi-\varphi_1(x), k(x))+F(\varphi_2(x), k(x)) - F(\pi-\varphi_2(x), k(x))
		2F(\varphi_2(x), k(x))-F(\varphi_1(x), k(x))
		& 0<x<1
		\\
		F(\varphi_1(x), k(x)), & x<0
	\end{cases}
\end{equation}
where modulus and angles are
\begin{align}
	&k(x) = \sqrt{\frac{1+|x|}{2|x|}},\quad \varphi_1=\arctan(\sqrt{|x|}),\quad \varphi_2=\arctan\sqrt{\frac{2|x|}{1-|x|}}.
\end{align}

%Since for $\abs{x}<1$ the modulus $k(x)>1$, we can transform elliptic integrals to modulus $k^{-1}<1$ if needed via the reciprocal-modulus transformation
%%
%\begin{equation}
%	F(k,\varphi) = \frac{1}{k}F\left(\frac{1}{k}, \beta\right), \quad \sin\beta=\frac{1}{k}\sin\varphi.
%\end{equation}

%\begin{figure}[h!]
%		\center{\includegraphics[width=0.7\linewidth]{DOS_supp_1}}
%	\caption{\textit{Left}: total DOS. \textit{Right}: DOS of each band }
%\end{figure}

\begin{figure}[t!]
	\begin{minipage}[h]{0.46\linewidth}
		\center{\includegraphics[width=0.99\linewidth]{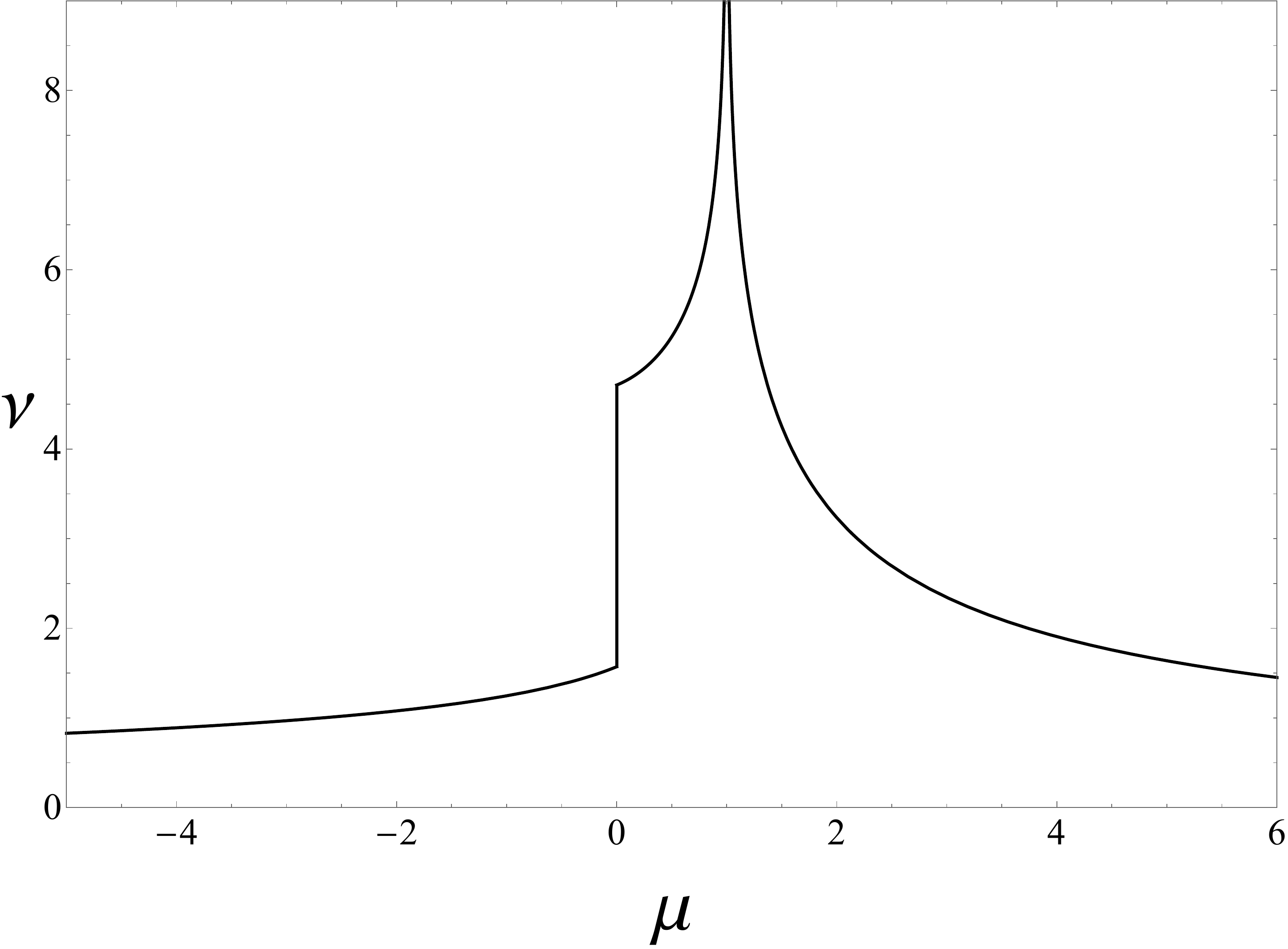}}
	\end{minipage}
	\hfill
	\begin{minipage}[h]{0.49\linewidth}
		\center{\includegraphics[width=0.99\linewidth]{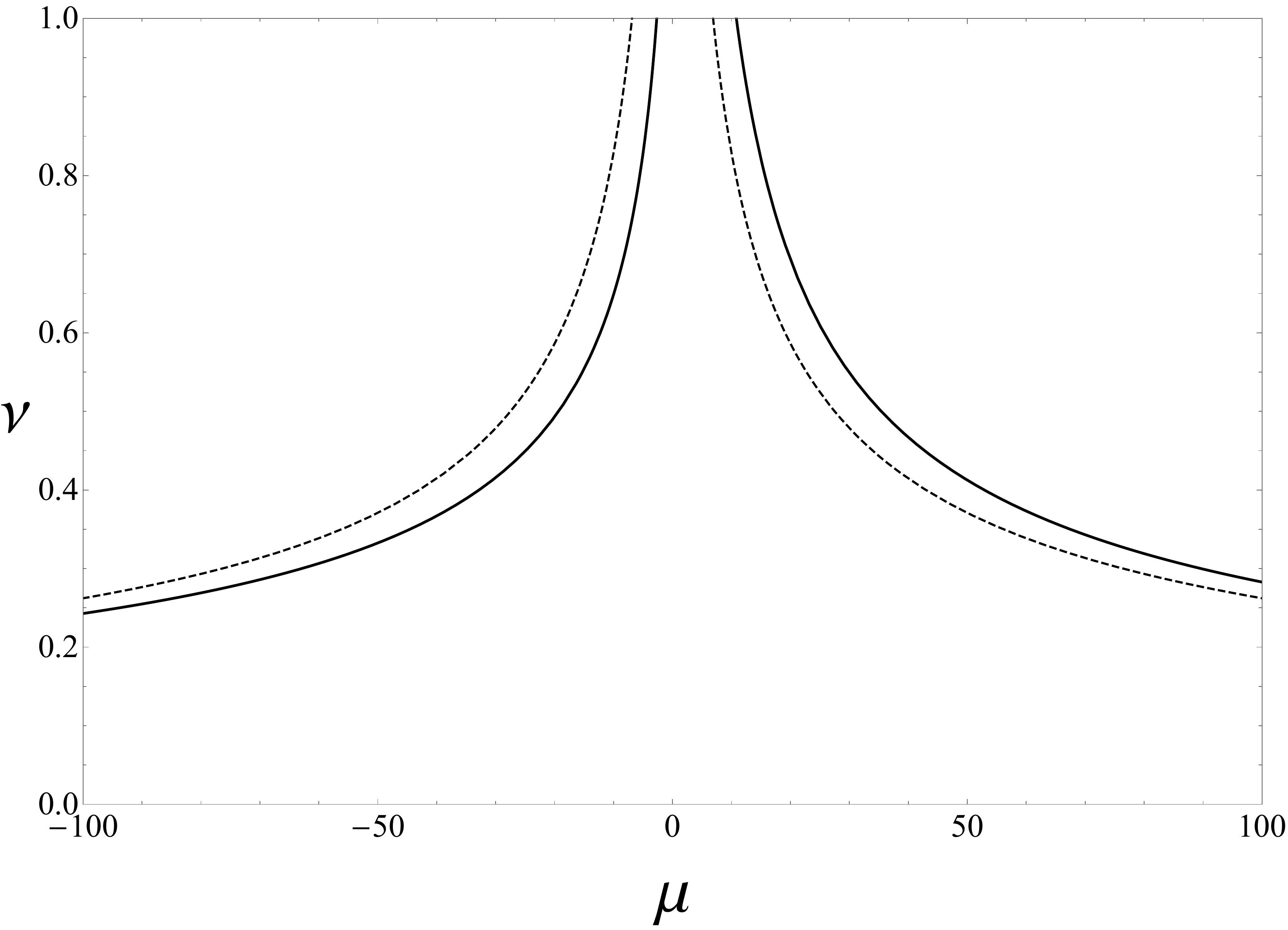}}
	\end{minipage}
	\caption{DoS $\nu(\mu)\equiv D(\mu)$, with DoS and chemical potential in units $(4\pi^2a)^{-1}$ and $\mu_c=a^2/4b$ respectively. The plot on the right shows DoS at larger values of the chemical potential when the singularity breaking term $ap^2$ can be neglected. The dashed line shows the critical scaling of the density of states given by (\ref{app:asymp}), $\nu(\mu)\propto{|\mu|}^{-1/2}$.}
\end{figure}

At large values of the argument the asymptotic behavior is
\begin{equation}
\label{app:asymp}
D(x)\simeq\sqrt{\frac{2}{|x|}}K\left(\frac{1}{\sqrt{2}}\right),\quad |x|\gg1,
\end{equation}

%Now we can confirm asymptotic behaviors derived above. At large values $\abs{x}\gg1$ $k(x)\rightarrow1/2$, $\varphi_1, (\pi-\varphi_1)\rightarrow \pi/2$, so
%%
%\begin{equation}
%\label{app:asymp}
%	D(x)\simeq\sqrt{\frac{2}{\abs{x}}}K\left(\frac{1}{\sqrt{2}}\right),\quad\abs{x}\gg1,
%\end{equation}
%%
%where $K(1/\sqrt{2})\approx1.85$ is a complete elliptic integral of the first kind.
%
%\newpage
%In the vicinity of $x=1$
%
%\begin{align}
%	D(1-\xi)\simeq D(1+\xi)\simeq \sqrt{2}F\left(\pi,1-\frac{\xi}{2}\right)\simeq 2^{3/2}K\left(1-\frac{\xi}{2}\right)\simeq\sqrt{2}\ln\frac{16}{\xi}.
%\end{align}
%
%Finally, around $x=0$
%\begin{align}
%D(-\xi)\simeq  \sqrt{\frac{2}{\xi}}F\left(\sqrt{\xi},k(-\xi)\right)\simeq \sqrt{\frac{2}{\xi}}\cdot\sqrt{\xi}\simeq\sqrt{2}+O(\xi\ln \xi).
%\end{align}

\section{Details of the  main features and the calculations relevant to Sr$_3$Ru$_2$O$_7$.}

\subsection{Details of the main features}

The magnetic susceptibility is strongly enhanced (Wilson ratio of 10), consistent with Sr$_3$Ru$_2$O$_7$ being on the border of ferromagnetism \cite{Ikeda}, achievable by modest uniaxial pressure \cite{Yaguchi}. 
Due to the observed anisotropic conductivities these have been referred to as ``electron nematics'' \cite{Borzi}.
The observed $Q$-vectors of the SDW (first discussed in \cite{Berridge}) might be related to FS nesting including the $\gamma$ band centered at the X-point, central to the theory developed below. Surprisingly, the entropy of the magnetically ordered A phase is higher than that of the adjacent low field phase \cite{Rost}.

The electronic band structure in the low field Fermi Liquid has been extensively studied experimentally \cite{Mercure, Tamai, Lee}. The layered crystal structure gives rise to an effectively 2D FS. In Fig.1e of the main text, we show schematically the key features both at the Fermi energy $E_F$  as well as at a slightly lower energy $E_{LT}$. At the latter there is an $n=4$ LT point occurring at the X-point of the Brillouin zone as becomes apparent by the characteristic clover leave structure in the extended zone (compare to Fig.1c of the main text). 

\subsection{Local (spin) density approximation.}

Our calculations were carried out within the local (spin) density approximation [L(S)DA] using the Full Potential Local Orbital band-structure package (FPLO) \cite{Koepernik, URL} based on structural data published in \cite{Shaked}. A mesh of 24 $\times$ 24 $\times$ 24 k-points in the whole Brillouin zone was employed. Due to the rather sizable spin-orbit interaction of the Ru atoms, the full relativistic four-component Dirac scheme was used. To test the generality of the results, we used two different functionals, local density approximation (LDA) and generalized gradient approximation (GGA).  The results are extremely similar for LDA and GGA, especially close to the Fermi energy.

\subsection{Effective Hamiltonian}

\begin{figure}[t]
	\begin{minipage}[h]{0.36\linewidth}%0.4 and 0.55 seem to give right scaling
		\center{\includegraphics[width=0.99\linewidth]{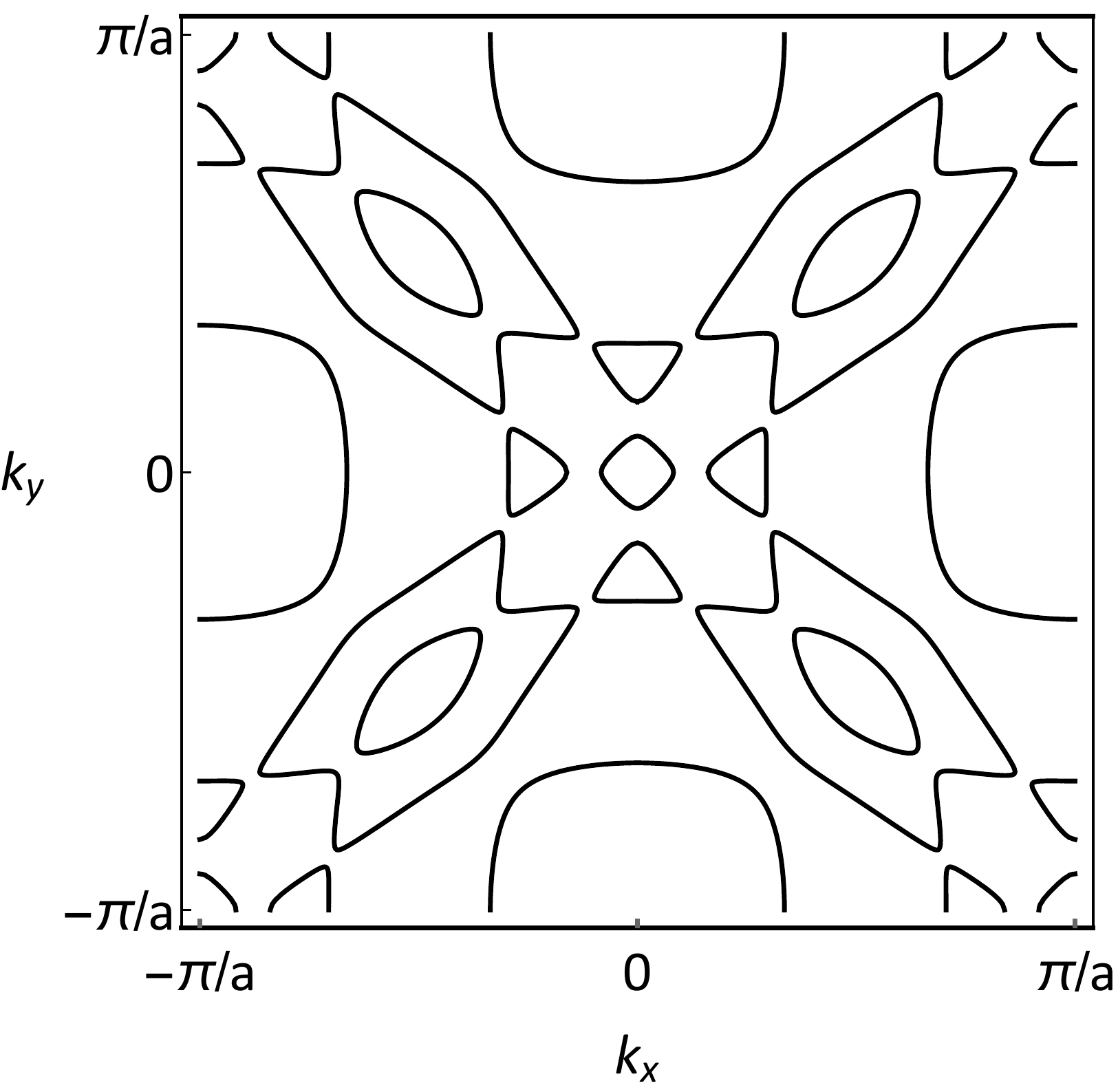}}
	\end{minipage}
	\hfill
	\begin{minipage}[h]{0.49\linewidth}
		\center{\includegraphics[width=0.99\linewidth]{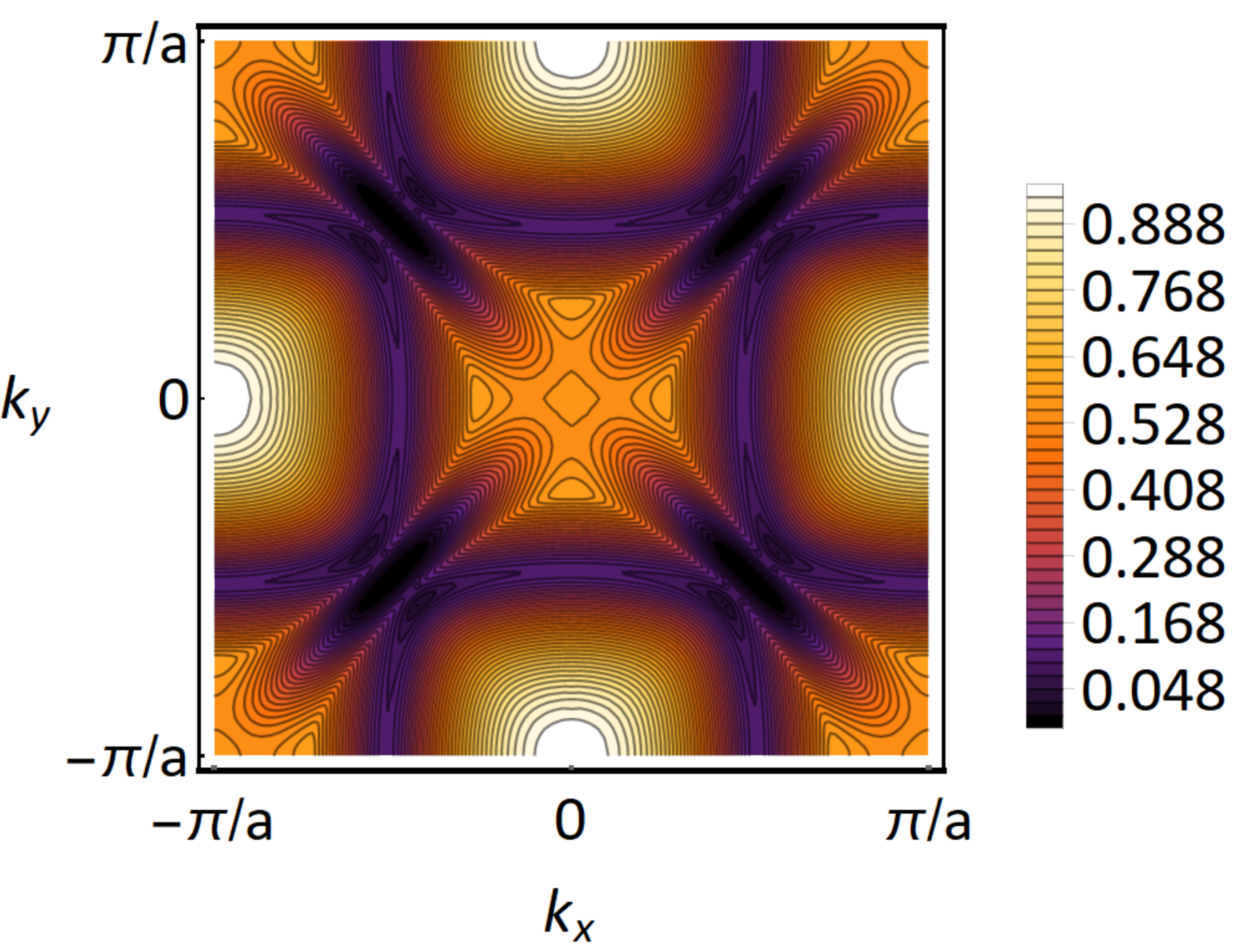}}
	\end{minipage}
	\caption{Fermi surface (left) and energy dispersion of the 4th band (right). The momentum is centered around $\gamma$ bands, where the Lifshitz multipoint resides. However, for given Hamiltonian parameters it is decomposed into four van Hove saddles}
	\label{app:figFS}
\end{figure}

In the main text we showed the presence of multicritical Lifshitz point in Sr$_3$Ru$_2$O$_7$ via DFT calculation. Here, we show that it is also present within a tight-binding model as presented in Ref. \onlinecite{Puetter}, which reads:
\begin{equation}
H=
\sum_{\vec{k},\alpha}\Psi^\dagger_{\vec{k}\alpha}
\begin{pmatrix}
	A_{\vec{k}} & G
	\\
	G^* & A_{\vec{k}+\vec{Q}}
\end{pmatrix}
\Psi_{\vec{k}\alpha},
\end{equation}
where $\alpha=\uparrow\downarrow$ is a spin index, $\vec{Q}=(\pi/a,\pi/a)$ and
\begin{equation}
A_{\vec{k}}  =
\begin{pmatrix}
\varepsilon_{\vec{k}}^{yz} & \varepsilon_{\vec{k}}^{1D} + i\lambda & -\lambda
\\
\varepsilon_{\vec{k}}^{1D} - i\lambda & \varepsilon_{\vec{k}}^{xz} & i\lambda
\\
-\lambda & i\lambda & \varepsilon_{\vec{k}}^{xy}
\end{pmatrix},
\quad G = 
\begin{pmatrix}
\bar{g} & 0 & 0
\\
0 & \bar{g} & 0
\\
0 & 0 & \bar{g}
\end{pmatrix}.
\end{equation}
The matrix $A$ describes the hopping between the three $yz, xz, xy$ orbitals with the spin-orbital coupling $\lambda$ included. The dispersions of the three individual orbitals are
\begin{equation}
\label{app:Puetter}
	\begin{split}
		\varepsilon_{\vec{k}}^{yz} &= -2t_1\cos(k_ya)-2t_2\cos(k_za),
		\\
		\varepsilon_{\vec{k}}^{xz} &= -2t_1\cos(k_xa)-2t_2\cos(k_za),
		\\
		\varepsilon_{\vec{k}}^{xy} &= -2t_3(\cos(k_xa)+\cos(k_ya)),
	\end{split}
\end{equation}
while $\varepsilon_{\vec{k}}^{1D}=-4t_6\sin(k_xa)\sin(k_ya)$ describes the hopping between the two quasi-one-dimensional orbitals $yz$ and $xz$. The matrix $G$ describes the unit-cell doubling via an effective lattice potential. The parameters of the Hamiltonian are
\begin{equation}
	t_1=0.5,\,t_2=0.05,\,t_3=0.5,\, t_4=0.1,\, t_5=-0.03,\, t_6=0.05,\, \lambda=0.1375,\, \bar{g}=0.1.
\end{equation}
The chemical potential can be determined self-consistently via the DoS that follows from (\ref{app:Puetter}) and is $\mu = 0.575$. Fermi surface is shown in the Fig.~\ref{app:figFS} (left), which is centered around $\gamma$ bands. For given parameters Hamiltonian is only in the vicinity of the topological multiple point and the singularity is decomposed into four van Hove saddles. 
The multicritical LT effectively originates from (i) 2D hopping terms on a square lattice (predominantly within the $d_{xy}$ orbitals) in combination with (ii) the doubling of the unit cell due to RuO-octahedra rotation (which also mixes in $e_g$ orbitals as a secondary effect) and (iii) bilayer splitting. 

Finally, even in this simple model there is a nesting between the parts of the $\gamma$ bands, that leads to the formation of the spin density wave. Examining eigenvalues of the Hamiltonian, we find that it is the 4th band that is involved in the formation of the multiple point around the Fermi energy. Given the $C_4$ rotational symmetry of the system, to show the presence of nesting, it is enough to calculate the curvature of this 4th band,
\begin{equation}
\kappa\left(k_x,k_y\right)=\frac{2E_xE_yE_{xy}-E_x^2E_{yy}-E_y^2E_{xx}}{(E_x^2+E_y^2)^{3/2}}
\end{equation}
where $E\equiv E(k_x,k_y)$ is the dispersion of the 4th band and $E_i\equiv \partial E/\partial k_i$ and $E_{ij}\equiv \partial^2 E/\partial k_i\partial k_j$ are partial derivatives. Using symmetry arguments again, it is enough to find whether the following condition can be satisfied:
\begin{equation}
\kappa\left(\frac{\vec{Q}}{2},0\right)\propto\frac{\partial^2E}{\partial k_y^2}\left(\frac{\vec{Q}}{2},0\right)=0.
\end{equation}
We calculate numerically the second derivative $E_{yy}$ of the electron dispersion of the 4th band and show that it turns to zero at the nesting vector $\vec{Q}=(\pm 0.29,0,0)(2\pi/a)$ (See Fig.~\ref{app:figNesting}). In this way we confirm the presence of the nesting in the system. The value we find is slightly different from the experimental finding of $\vec{Q}_\text{exp}=(\pm 0.233,0,0)(2\pi/a)$. It is important to stress that
\begin{itemize}
	\item it is a simple tight-binding model without interactions.
	\item there is no fine-tuning of the effective Hamiltonian, as exactly the same parameters as in Ref. \onlinecite{Puetter} has been used.
\end{itemize}
%
%One can indeed modify the parameters of the system as to reproduce the singularity and the value of the nesting vector. However, the validity of such endeavor is very much questionable. 
As a result, here we confirm the presence of the nesting using the tight binding Hamiltonian, while the actual quantitative features are reproduced with the \textit{ab initio} approach of LDA calculation presented in the main text.

\begin{figure}[h]
	\center{\includegraphics[width=0.75\linewidth]{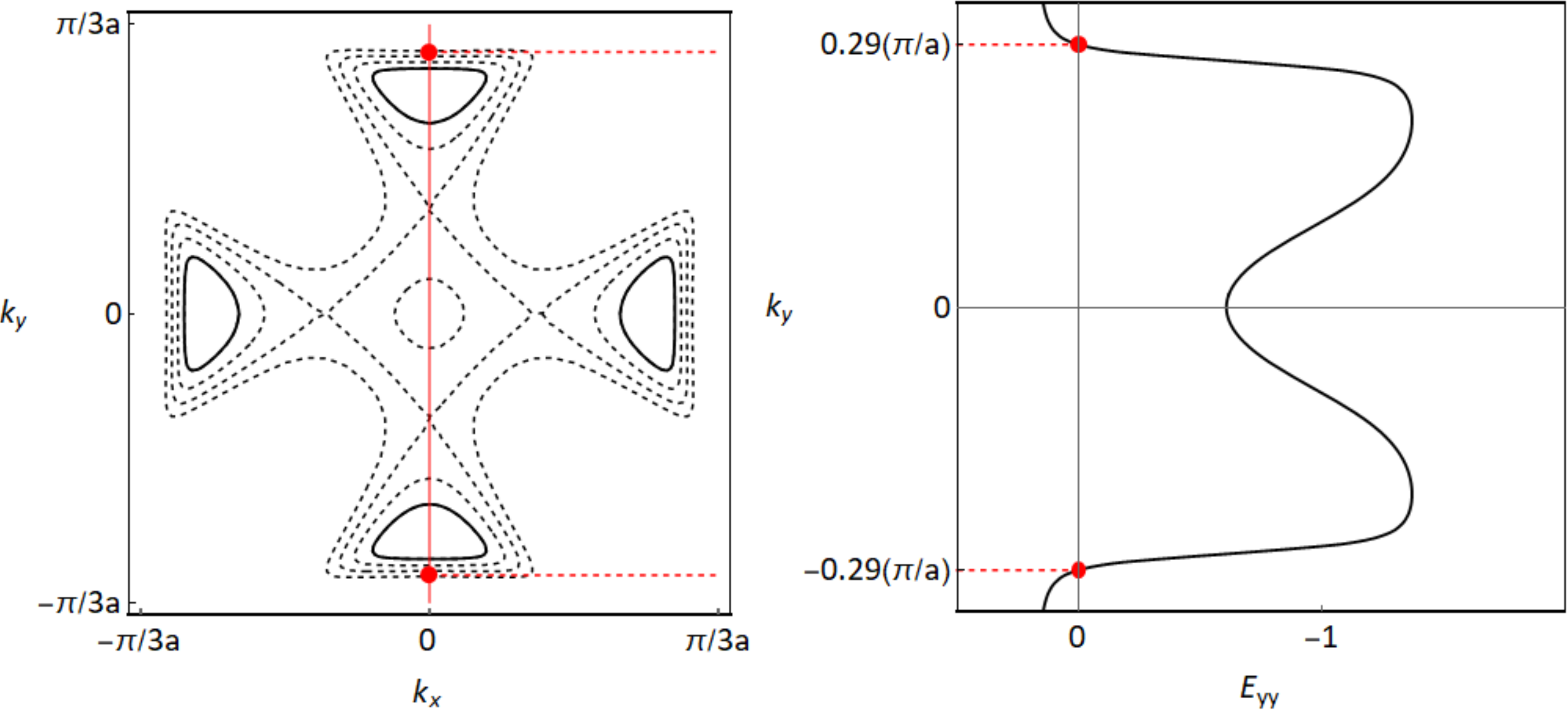}}
	\caption{Surfaces (left) of constant energy for the 4th band and the second derivative $E_{yy}$ along the $y$-axis (right). The second derivative $E_{yy}$ turns to zero at $\vec{Q}/2=(0.29,0,0)(\pi/a)$ indicating a flat Fermi surface at this energy and the presence of the nesting.}
	\label{app:figNesting}
\end{figure}

\subsection{Confirmation of the dispersion with ARPES data}

\begin{figure}[h]
\begin{center}
\includegraphics[width=0.95\textwidth]{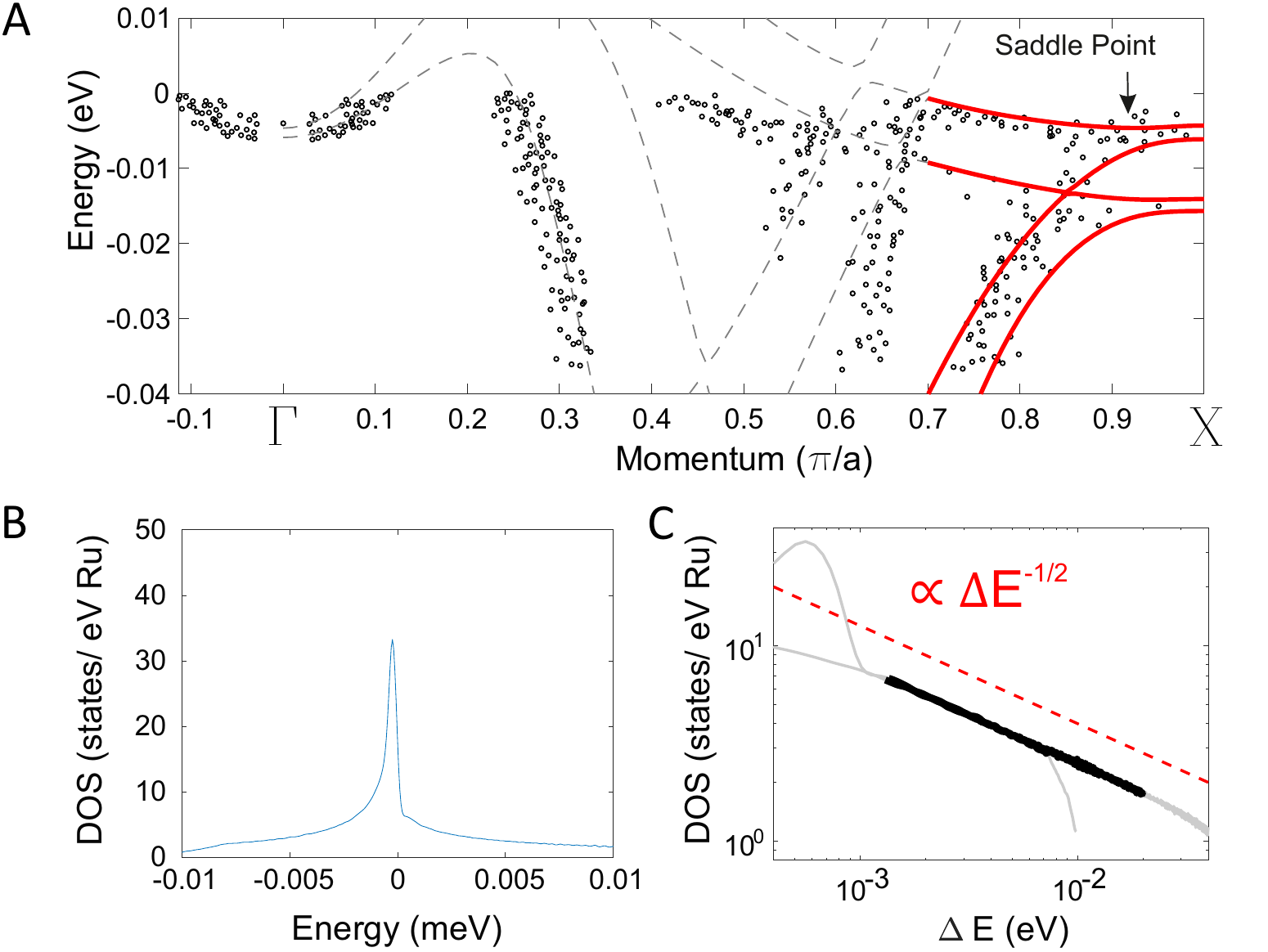}
\end{center}
\caption{\label{fig:ARPES} Divergence in density of states based on ARPES dispersion. For Details see text.}
\end{figure}

In this subsection we present the best possible analysis based on available high resolution ARPES data in combination with the band structure topology known from DFT/tight binding models.
In order to determine the regime of parameters for Sr$_3$Ru$_2$O$_7$ where the expansion provides accurate results, we start from a tight binding model based on the DFT calculation and carry out a minimal orbital dependent renormalisation of the band width and chemical potential (Ref. (\onlinecite{Allan})), matching the available high resolution experimental (ARPES) data of the band structure (with sufficient energy resolution this data has only been published along the  $\Gamma$-X direction in Ref. (\onlinecite{Allan}). In Fig.~\ref{fig:ARPES} we show the data from Ref. (\onlinecite{Allan}) digitised by us together with the adjusted band structure of the tight binding model (dotted lines). In red we emphasise the relevant $d_{xy}$ bands in the crucial range of momentum space with the saddle point location marked by an arrow. We would like to point out that in our tight binding model only a limited number of hopping terms can be included accounting for some of the remaining discrepancies. 
We fitted the expansion of Eq. (1) of the manuscript (incl. higher order) terms to the relevant band of the tight binding model (uppermost red band in panel (A)) and find the crucial parameter K to be -2.8(1) with the band structure being given by (terms in bracket give uncertainty)

\begin{equation}
E (eV)= 0.131(2) \left[-0.033(1)-0.088(1)(k_x^2+k_y^2 )-2.8(1) k_x^2 k_y^2 +(k_x^4+k_y^4 )\right]+O(k_x^6,k_y^6)
\end{equation}

In panel (B) and (C) we present the density of states of this part of the band structure both on a linear scale as well as on a log-log scale ($\Delta E$ is measured relative to $\mu_c$ and an offset added to account for the background density of states). In particular in panel (C) we show the total density of states in grey. We highlight in black the energy regime over which the density of states is well described by a power law with critical exponent -1/2 (as is shown by the parallel red dotted line). Most crucially the power-law like divergence covers both the Fermi energy as well as the experimentally relevant energy scales.

Given the experimental uncertainties this is the most accurate possible description of the band structure in the vicinity of the multicritical Lifshitz point and implies that the density of states is well described by a power-law divergences over several meV around the singularity. 

\section{Entropy near $H_c$ for an SDW instability}
\label{appendix3}

According to scaling theory, for the SDW transition \cite{Zhu}, if the critical part of the free energy is separated, then ${\cal{F}}_{crit} ^{SDW}= -\rho_0 (\frac{T}{T_0})^{\frac{d+z}{z}} f((H-H_c)/(T/T_0)^{\frac{1}{\nu z}}) $ + logarithmic corrections as $\frac{T_0 (H-H_c)}{TH_c} \rightarrow 0$. $T_0$ and $\rho_0$ and $c$ are non-universal constants, $d=2$ is the dimensionality of the system, $z=2$ the dynamical exponent, $\nu=1/2$ the exponent that shows the divergence of the coherence length $\xi$ close to the transition $\xi \propto |\frac{H-H_c}{H_c}|^{-\nu}$ and $f(x)$ is a regular function, which can be expanded close to $x=0$. Therefore in the limit $\frac{T_0(H-H_c)}{TH_c} \rightarrow 0$
\begin{equation}
{\cal{F}}_{crit}^{SDW} = -\rho_0 \left[ \left( \frac{T}{T_0} \right)^2 f(0) +  \frac{H-H_c}{H_c} \frac{T_0}{T}  f^{\prime}(0) + c log\left(\frac{H_c}{|H-H_c|}\right) \right]
\end{equation}
\noindent leading to an increase in entropy close to $H_c$ due to the logarithmic divergence.

%When the system is tuned precisely into the critical chemical potential, specific heat scales as
%%
%\begin{equation}
%C(T,\mu=0)\propto
%\begin{cases}
%\left|T\right|^{-1/2}, & T\gg a^2
%\\
%\ln \dfrac{a^2}{T}, & \left|T\right|\ll a^2
%\end{cases}.
%\end{equation}

\section{Higher order singularities in other materials}
\label{appendix4}

The pivotal concept controlling the wealth of phenomena observed in Sr$_3$Ru$_2$O$_7$ is the existence of a multicritical Lifshitz point interacting with an incipient nesting feature. Naturally a key question is the generality of this concept and its suitability as a guiding principle in material design. 

The aim of this section is to show that there are indeed generic mechanisms generating such multicritical Lifshitz points in a wide range of material classes making them an extremely valuable 'material design guide'. A detailed analysis of all materials discussed here along the lines of the one presented for Sr$_3$Ru$_2$O$_7$ would certainly go beyond the scope of this paper. However, we believe that the existence of multicritical Lifshitz points in the material classes identified below play an important role in the boosting and stabilising the observed physical phenomena and should be taken into consideration in the analysis of effective low energy hamiltonians of these systems.

\subsection{Direct generalisation of Sr$_3$Ru$_2$O$_7$}

In order to appreciate the generality of the multicritical Lifshitz point in Sr$_3$Ru$_2$O$_7$ it is instructive to scrutinize the details of a low energy tight binding model based on the relevant Ru 4$d$ orbitals. In figure \ref{fig:SDisp}a we show the DFT band structure calculation described in the main text along the relevant high symmetry directions in the $k_z=0$ plane of the 3D Brillouin zone of this quasi-2D material. In red we emphasise the part of the band structure giving rise to the multicritical Lifshitz point. Following previous work \cite{Fischer, Piefke, Autierie, Rozbicki} we developed an effective 2D tight binding model of a bilayer of RuO octahedra constrained to the relevant Ru 4$d$ orbitals. In figure \ref{fig:SDisp}b we show the dispersion of this tight binding model along the equivalent high symmetry directions of figure \ref{fig:SDisp}a with an overall good agreement.  

\begin{figure}[b!]
\begin{center}
\includegraphics[width=0.9\textwidth]{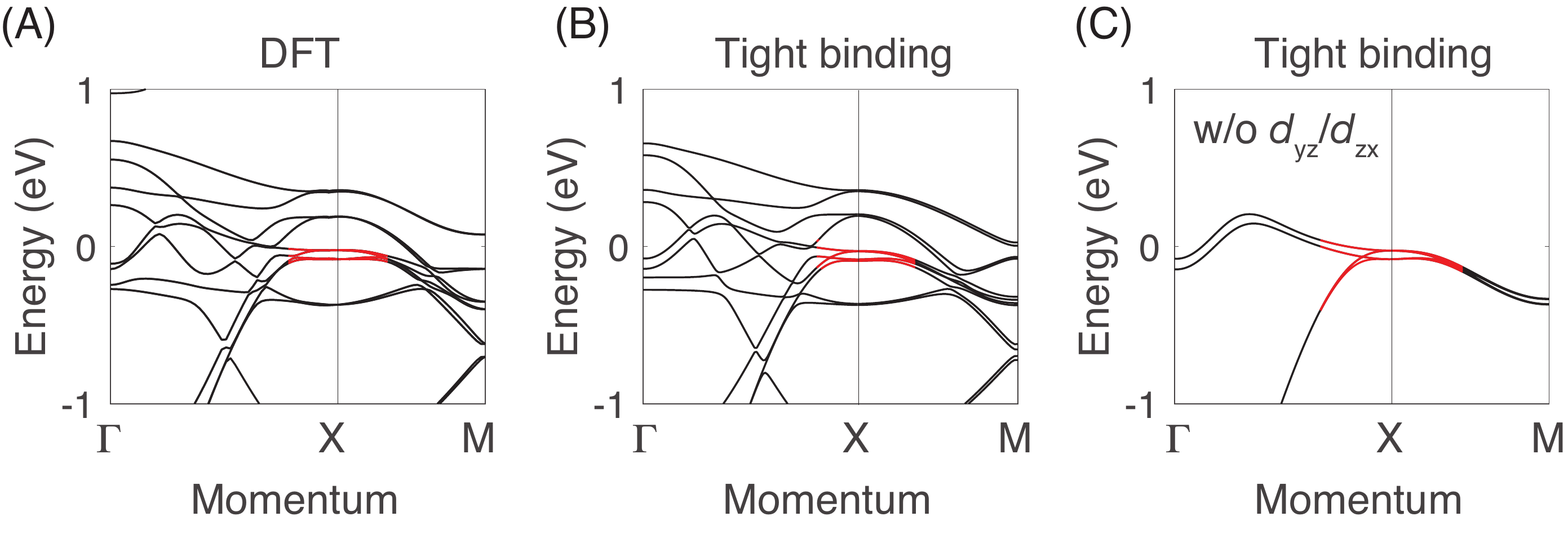}
\end{center}
\caption{\label{fig:SDisp}(a) Result of the DFT calculation along relevant high symmetry directions of the Brillouin zone. The part of the band structure giving rise to the multicritical Lifshitz point is emphasised in red. (b) Dispersion of an effective 2D tight binding model of a bilayer of RuO octahedra based on Ru 4$d$ orbitals. (c) Same tight binding model as in (b) but with the Ru 4$d_{yz/zx}$ orbitals removed from the model.}
\end{figure}

In this energy regime the orbital character of the band structure is predominantly that of the $t_{2g}$ submanifold containing $d_{xy}$, $d_{yz}$ and $d_{zx}$ orbitals.One of the advantages of the tight binding model is the possibility of selectively removing orbitals and / or in order to study the impact on the relevant band structure features. In figure \ref{fig:SDisp}c we have for example removed the $d_{yz}$ and $d_{zx}$ orbitals. As can be seen this effectively leaves the dispersion giving rise to the multicritical Lifshitz point unchanged. This of course is equivalent to the observation that in the DFT calculation the relevant part of the band structure has effectively no 4$d_{yz/zx}$ character and is predominantly $d_{xy}$ with a small contribution of the $e_g$ orbitals \cite{Rozbicki}. 

At this point it is important to briefly discuss in how far the DFT band structure is relevant to the experimentally observed band structure of this strongly correlated material. Direct observation of the Fermi surface and the determination of effective masses and Fermi velocities has been achieved by ARPES\cite{Tamai, Allan}, STM\cite{Lee} and quantum oscillation\cite{Mercure} measurements. Crucially in particular the ARPES work has demonstrated that the band structure in the vicinity of the X-point is well described by an overall renormalised band structure and is qualitatively well described by LDA. We have verified that the tight binding model described below is fully consistent with the van Hove singularity observed in ARPES\cite{Tamai} confirming the arguments for generality presented here.

The two copies of the band structure apparent in \ref{fig:SDisp}c are the result of a bilayer split. Otherwise the tight binding model is that of a 2D square lattice with equivalent hopping along $x$ and $y$ axis including the following features:
\begin{enumerate}
\item A $\sqrt 2$ x $\sqrt 2$ reconstruction of the Brillouin zone. In the case of Sr$_3$Ru$_2$O$_7$ this is achieved by a counterrotation of neighbouring octahedra but could also be due to e.g. antiferromagnetic ordering or charge disproportionation or the momentum space picture of $(\pi,\pi)$ charge-/spin-density-waves.
\item a small hybridisation gap opening at the zone boundary and in particular at the high-symmetry $X$ point between the original and backfolded bands.
\end{enumerate}

These ingredients are fully sufficient to generate the multicritical Lifshitz point. The resulting band structure (constrained by ARPES \cite{Tamai}) in the vicinity of the $X$-point retaining these minimal ingredients is shown in figure \ref{fig:MinMod}. In panel (a) we reproduce a schematic of the Fermi surface of Sr$_3$Ru$_2$O$_7$ for orientation (based on \cite{Mercure}). The multicritical Lifshitz point is located close to the X-point. In red we highlight the same parts of the band structure as in figure \ref{fig:SDisp}. The inset in the top-right corner shows the location and orientation of the contour plots shown in panel (b) and (c). In the former we present the tight binding model excluding the 4$d_{yz/zx}$. The location of the Lifshitz points is indicate by the red crossed. Intriguingly the Fermi surface of this model does not produce the small closed pockets highlighted in panel (a). These are only recovered if the much more strongly dispersing bands originating from 4$d_{yz/zx}$ are reintroduced  as shown in panel (c) \footnote{Qualitative similar Hamiltonians have been previously postulated (e.g. \onlinecite{Berridge2}). Here we have followed a more rigorous line of argument starting from the DFT calculation and tight binding model allowing a more precise determination of orbital character and origin of the observed band structure. However qualitatively the Hamiltonians are equivalent.}. Remarkably it is the resulting flat parts that give rise to a nesting vector (blue) consistent with the spin density wave observed in neutron scattering \cite{Lester}. It is this feature that drives the density wave phase within the theoretical model discussed in the main text. Within our model it is merely coincidence that the nested part of the band structure determining the density wave vector is part of the same Fermi surface giving rise to the multicritical Lifshitz point - these two components originate from completely separate, weakly hybridising bands with very different orbital character. From a material design point of view this is extremely important in that a multicritical Lifshitz point can boost / thermodynamically stabilise incipient order driven by otherwise unrelated parts of the Fermi surface.  

\begin{figure}[t!]
\begin{center}
\includegraphics[width=0.9\textwidth]{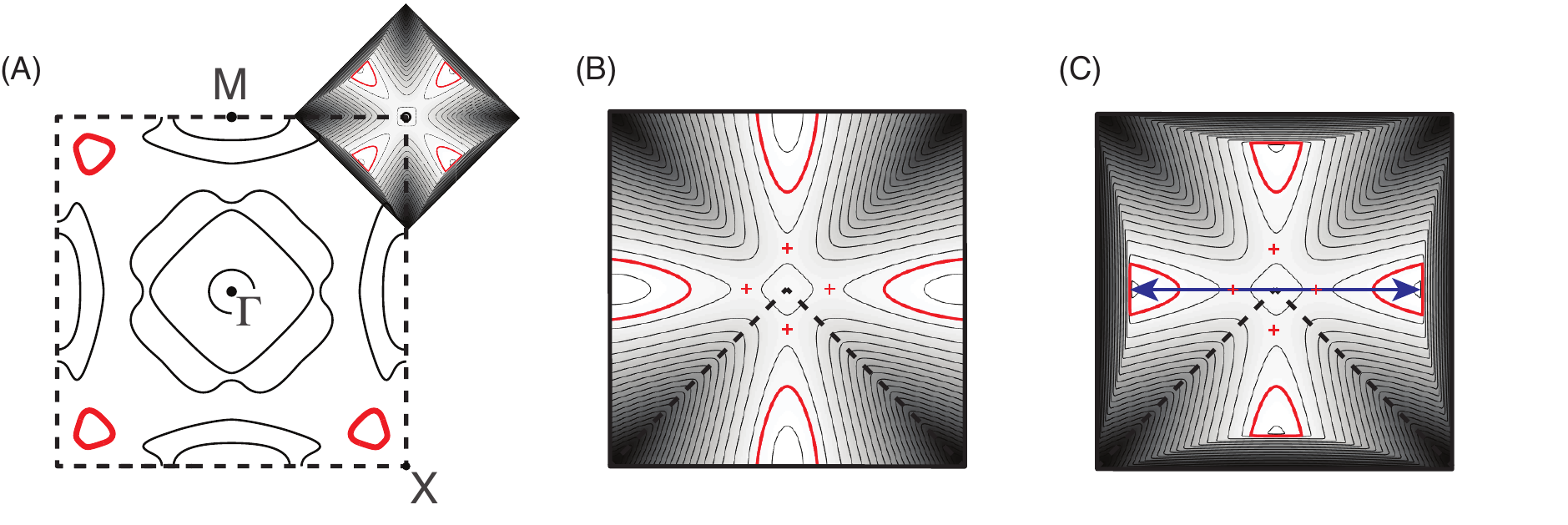}
\end{center}
\caption{\label{fig:MinMod} (a) Schematic of Fermi surface of Sr$_3$Ru$_2$O$_7$ (based on a figure in \cite{Mercure}). The red parts of the Fermi surface are those relevant to the multicritical Lifshitz point. they are generally referred to as $\gamma_2$ pockets in the literature. The overlay centred on the top right X-point is showing the location of the contour plots in the subsequent panels. (b) Contour plot of the minimal model generating multicritical Lifshitz points (red crosses). (c) Upon including the $d_{yz/zx}$ orbitals the $\gamma_2$ pockets are reproduced. This results in nested Fermi surface features (blue arrow) originating from different bands than the multicritical Lifshitz point (red cross).}
\end{figure}

As a side note we would like to remark that a fit of equation (3) in the main text to this low energy band structure results in $K$ being of the order of -3 consistent with our analysis. Indeed it can be shown that a hamiltonian of form (B1) in ref.  \cite{Berridge} will generally result in $K\leq-2$ as required for the validity of our theoretical analysis.

Irrespective of the detailed parameters for Sr$_3$Ru$_2$O$_7$ it is remarkable that such few ingredients are required for the stabilisation of the multicritical Lifshitz point for a tight binding model on a square lattice. Consequently one would expect these to be e.g. ubiquitous across the whole family of ruthenates. Indeed further examples in this family are:

\textbf{\textit{Ruthenates - Surface layer of Sr$_2$RuO$_4$}}
The most straight forward realisation of a multicritical Lifshitz point as described above is occurring in the surface layer of Sr$_2$RuO$_4$. All key ingredients exist: (i) a vHs at the M point of the ideal band structure, (ii) backfolding due to rotation of RuO$_2$ octahedra changing the C$_2$ symmetric singularity into a C$_4$ symmetric one at the X-point of the new surface Brillouin zone  and finally (iii) hybridisation of the backfolded bands due to higher order hopping terms. Indeed ARPES measurements reveal a band structure consistent with this model located just 10meV below the Fermi energy\cite{Rozbicki}. It would be interesting to study the evolution of the surface electronic structure in spectroscopic studies under magnetic field by e.g. STM.

\textbf{\textit{Ruthenates - Ca$_3$Ru$_2$O$_7$}}  In the case of Ca$_3$Ru$_2$O$_7$ the relevant physics originates not from a $d_{xy}$ band but from two $d_{xz/yz}$ bands with next nearest neighbour hopping \cite{Kikugawa}. An alternating $c$-axis tilt of the RuO octahedra then generates a backfolding of the band structure. At this point there is no singularity in the density of states at half filling. However, the combination of bilayer splitting and hybridisation generate singularities in the density of states located at the $X$-points of the Brillouin zone. Most interestingly this system subsequently undergoes a density wave order transition at app. 40 K \cite{Baumberger}. 

\textbf{\textit{Cuprates}} The nature of charge- and spin-excitations as well as (short range) density wave formation in cuprates is of course a subject of ongoing discussions. However an often used starting point for theoretical calculations that is well motivated by experimental observations by ARPES and quantum oscillations is a band structure originating from the close-to-half filled $d_{xy}$ band on a square lattice. This of course is exactly the situation as encountered in Sr$_3$Ru$_2$O$_7$. Consequently any $\sqrt 2$ x $\sqrt 2$ reconstruction of the unit cell (by e.g. (short range) magnetic order) in combination with a hybridisation gap opening between the original and back-folded bands will therefore generate exactly the same type of multicritical Lifshitz point as discussed for the example of Sr$_3$Ru$_2$O$_7$. This by no means is intended to be an explanation for the wealth of strong-coupling phenomena observed in cuprates but is merely to be seen as an exemplification of the generality of the concepts discussed in this paper and should potentially be taken carefully into account when discussing effective weak coupling hamiltonians.

\subsection{Further materials}
In this final section we would like to highlight the breadth of materials across many different 'families' in which multicritical Lifshitz points occur close or at the Fermi energy with potentially crucial impact on the stabilisation of incipient ordered phases. We would like to emphasise that neither is this an exhaustive list nor are we giving a review over the literature on these materials but merely intend to provide the interested reader with starting points to the extensive experimental literature.

\textbf{\textit{BaFe$_2$As$_2$-Iron based superconductors}} In figure \ref{fig:SDisp}a we show schematically the Fermi surface of BaFe$_2$As$_2$ based superconductors (see e.g. \cite{Borisenko} for a more recent paper and citations therein). The typical 'clover leaf' structure of a Fermi surface close to a multicritical Lifshitz point can be clearly identified close to the X-Point. Recently it has been shown that this part of the band structure gives rise to a singular Fermi surface in the related compound SmFe$_0.92$Co$_0.08$AsO \cite{Borisenko}.

\textbf{\textit{Transition metal dichalcogenides - 1T-VSe$_2$}} In figure \ref{fig:SDisp}b we show the evolution of the band structure with energy of a monolayer of the transition metal dichalcogenide VSe$_2$ recently studied by ARPES \cite{King}. This is an example of an $n=6$ multicritical Lifshitz point close to the $\Gamma$-point of the Brillouin zone. The singularity is located within 20\~meV of the Fermi surface. Intriguingly a charge density wave order setting in at $T_C$=140\~K has been observed in the same study. Similar physics has been seen in the related 3D band structure / phase diagram of bulk VSe$_2$ \cite{Tsutsumi, Eaglesham} and other transition metal dichalcogenides  (see e.g. references in the introduction of \cite{King}).

\textbf{\textit{Twisted Bilayer Graphene}} Recently twisted bilayer graphene has come to the fore of condensed matter research by the experimental discovery of interaction driven insulating behaviour and the development of unconventional superconductivity 
\cite{Cao1,Cao2}. The band structure of these materials gives naturally rise to $n=3$ multicritical points at the K/K' points of the 2D Brillouin zone (see figure \ref{fig:SDisp}b) and it can be shown that these play a potentially crucial role in the stabilisation of the observed electron-correlation driven phases \cite{Shtyk,Sherkunov}.

\begin{figure}[t!]
\begin{center}
\includegraphics[width=0.9\textwidth]{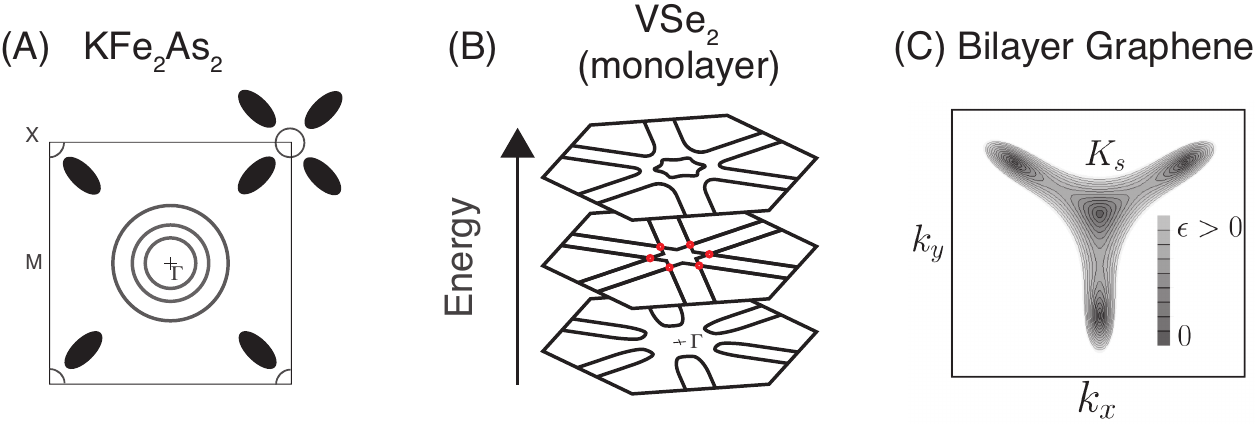}
\end{center}
\caption{\label{fig:SDisp}(a) Fermi surface of BaFe$_2$As$_2$ based iron based superconductors. in the top right corner we emphasise the typical 'glover leave' Fermi surface structure of incipient $n=4$ multicritical Lifshitz points (see e.g. \cite{Borisenko}). (b) Evolution of the band structure with energy of a monolayer of the transition metal dichalcogenide VSe$_2$  \cite{King}. The Lifshitz points are highlighted (red). (c) Contour plot of the $n=3$ multicritical Lifshitz point (monkey saddle) in twisted bilayer graphene (reproduced from \cite{Sherkunov}.}
\end{figure}

\end{widetext}
\end{document}